\documentclass[aps, prd,twocolumn,superscriptaddress,nofootinbib,preprintnumbers]{revtex4-1}
\usepackage[utf8]{inputenc}
\usepackage[english]{babel}
\usepackage{natbib}
\def\apj{ApJ}%
\def\apjl{ApJ}%
\def\aap{A\&A}%
\def\mnras{MNRAS}%
\def\prd{Phys.~Rev.~D}%
\def\pasp{PASP}%
\def\jcap{JCAP}%
\def\physrep{Phys.~Rep.}%
\usepackage{graphicx}
\usepackage{amsmath,amssymb}
\usepackage{hyperref}
\usepackage{braket}
\usepackage{subfigure}
\usepackage{float}
\usepackage[dvipsnames]{xcolor}
\usepackage{soul}
\usepackage{rotating}
\usepackage{multirow}
\usepackage{mathtools}
\usepackage{makecell}

\usepackage[margin=0.75in]{geometry}

\hypersetup{
	colorlinks=true,
	linkcolor=red,
	citecolor=blue,
}

\usepackage{xcolor}

\newcommand{\beq}{\begin{equation}}
\newcommand{\eeq}{\end{equation}}
\newcommand{\beqa}{\begin{eqnarray}}
\newcommand{\eeqa}{\end{eqnarray}}
\newcommand{\bk}{\mathbf{k}}
\definecolor{darkgreen}{rgb}{0.0, 0.5, 0.0}
\definecolor{darkcyanxf}{RGB}{0.0, 139.0, 139.0}

\newcommand{\redmagic}{\texttt{redMaGiC} }
\newcommand{\mice}{\texttt{MICE} }
\newcommand{\maglim}{\texttt{Maglim} }

\begin{document}
\title[Bias Validation]{\boldmath Perturbation theory for modeling galaxy bias: validation with simulations of the Dark Energy Survey}

% Author list file generated with: mkauthlist 1.2.4 
% mkauthlist -f -d -j prd /Users/shivam/Dropbox/Downloads/DES-2019-0458_author_list_ascii.csv DES-2019-0458_author_list.tex 

% \documentclass[reprint,superscriptaddress]{revtex4-1}
% \pagestyle{empty}
% \begin{document}
% \title{DES Publication Title}

\author{S.~Pandey}
\affiliation{Department of Physics and Astronomy, University of Pennsylvania, Philadelphia, PA 19104, USA}
\author{E.~Krause}
\affiliation{Department of Astronomy/Steward Observatory, University of Arizona, 933 North Cherry Avenue, Tucson, AZ 85721-0065, USA}
\author{B.~Jain}
\affiliation{Department of Physics and Astronomy, University of Pennsylvania, Philadelphia, PA 19104, USA}
\author{N.~MacCrann}
\affiliation{Center for Cosmology and Astro-Particle Physics, The Ohio State University, Columbus, OH 43210, USA}
\affiliation{Department of Physics, The Ohio State University, Columbus, OH 43210, USA}
\author{J.~Blazek}
\affiliation{Center for Cosmology and Astro-Particle Physics, The Ohio State University, Columbus, OH 43210, USA}
\affiliation{Institute of Physics, Laboratory of Astrophysics, \'Ecole Polytechnique F\'ed\'erale de Lausanne (EPFL), Observatoire de Sauverny, 1290 Versoix, Switzerland}
\author{M.~Crocce}
\affiliation{Institut d'Estudis Espacials de Catalunya (IEEC), 08034 Barcelona, Spain}
\affiliation{Institute of Space Sciences (ICE, CSIC),  Campus UAB, Carrer de Can Magrans, s/n,  08193 Barcelona, Spain}
\author{J.~DeRose}
\affiliation{Department of Astronomy, University of California, Berkeley,  501 Campbell Hall, Berkeley, CA 94720, USA}
\affiliation{Santa Cruz Institute for Particle Physics, Santa Cruz, CA 95064, USA}
\author{X.~Fang}
\affiliation{Department of Astronomy/Steward Observatory, University of Arizona, 933 North Cherry Avenue, Tucson, AZ 85721-0065, USA}
\affiliation{Department of Physics, University of Arizona, Tucson, AZ 85721, USA}
\author{I.~Ferrero}
\affiliation{Institute of Theoretical Astrophysics, University of Oslo, P.O. Box 1029 Blindern, NO-0315 Oslo, Norway}
\author{O.~Friedrich}
\affiliation{Kavli Institute for Cosmology, University of Cambridge, Madingley Road, Cambridge CB3 0HA, UK}
\author{M.~Aguena}
\affiliation{Departamento de F\'isica Matem\'atica, Instituto de F\'isica, Universidade de S\~ao Paulo, CP 66318, S\~ao Paulo, SP, 05314-970, Brazil}
\affiliation{Laborat\'orio Interinstitucional de e-Astronomia - LIneA, Rua Gal. Jos\'e Cristino 77, Rio de Janeiro, RJ - 20921-400, Brazil}
\author{S.~Allam}
\affiliation{Fermi National Accelerator Laboratory, P. O. Box 500, Batavia, IL 60510, USA}
\author{J.~Annis}
\affiliation{Fermi National Accelerator Laboratory, P. O. Box 500, Batavia, IL 60510, USA}
\author{S.~Avila}
\affiliation{Instituto de Fisica Teorica UAM/CSIC, Universidad Autonoma de Madrid, 28049 Madrid, Spain}
\author{G.~M.~Bernstein}
\affiliation{Department of Physics and Astronomy, University of Pennsylvania, Philadelphia, PA 19104, USA}
\author{D.~Brooks}
\affiliation{Department of Physics \& Astronomy, University College London, Gower Street, London, WC1E 6BT, UK}
\author{D.~L.~Burke}
\affiliation{Kavli Institute for Particle Astrophysics \& Cosmology, P. O. Box 2450, Stanford University, Stanford, CA 94305, USA}
\affiliation{SLAC National Accelerator Laboratory, Menlo Park, CA 94025, USA}
\author{A.~Carnero~Rosell}
\affiliation{Instituto de Astrofisica de Canarias, E-38205 La Laguna, Tenerife, Spain}
\affiliation{Universidad de La Laguna, Dpto. Astrofisica, E-38206 La Laguna, Tenerife, Spain}
\author{M.~Carrasco~Kind}
\affiliation{Department of Astronomy, University of Illinois at Urbana-Champaign, 1002 W. Green Street, Urbana, IL 61801, USA}
\affiliation{National Center for Supercomputing Applications, 1205 West Clark St., Urbana, IL 61801, USA}
\author{J.~Carretero}
\affiliation{Institut de F\'{\i}sica d'Altes Energies (IFAE), The Barcelona Institute of Science and Technology, Campus UAB, 08193 Bellaterra (Barcelona) Spain}
\author{M.~Costanzi}
\affiliation{INAF-Osservatorio Astronomico di Trieste, via G. B. Tiepolo 11, I-34143 Trieste, Italy}
\affiliation{Institute for Fundamental Physics of the Universe, Via Beirut 2, 34014 Trieste, Italy}
\author{L.~N.~da Costa}
\affiliation{Laborat\'orio Interinstitucional de e-Astronomia - LIneA, Rua Gal. Jos\'e Cristino 77, Rio de Janeiro, RJ - 20921-400, Brazil}
\affiliation{Observat\'orio Nacional, Rua Gal. Jos\'e Cristino 77, Rio de Janeiro, RJ - 20921-400, Brazil}
\author{J.~De~Vicente}
\affiliation{Centro de Investigaciones Energ\'eticas, Medioambientales y Tecnol\'ogicas (CIEMAT), Madrid, Spain}
\author{S.~Desai}
\affiliation{Department of Physics, IIT Hyderabad, Kandi, Telangana 502285, India}
\author{J.~Elvin-Poole}
\affiliation{Center for Cosmology and Astro-Particle Physics, The Ohio State University, Columbus, OH 43210, USA}
\affiliation{Department of Physics, The Ohio State University, Columbus, OH 43210, USA}
\author{S.~Everett}
\affiliation{Santa Cruz Institute for Particle Physics, Santa Cruz, CA 95064, USA}
\author{P.~Fosalba}
\affiliation{Institut d'Estudis Espacials de Catalunya (IEEC), 08034 Barcelona, Spain}
\affiliation{Institute of Space Sciences (ICE, CSIC),  Campus UAB, Carrer de Can Magrans, s/n,  08193 Barcelona, Spain}
\author{J.~Frieman}
\affiliation{Fermi National Accelerator Laboratory, P. O. Box 500, Batavia, IL 60510, USA}
\affiliation{Kavli Institute for Cosmological Physics, University of Chicago, Chicago, IL 60637, USA}
\author{J.~Garc\'ia-Bellido}
\affiliation{Instituto de Fisica Teorica UAM/CSIC, Universidad Autonoma de Madrid, 28049 Madrid, Spain}
\author{D.~Gruen}
\affiliation{Department of Physics, Stanford University, 382 Via Pueblo Mall, Stanford, CA 94305, USA}
\affiliation{Kavli Institute for Particle Astrophysics \& Cosmology, P. O. Box 2450, Stanford University, Stanford, CA 94305, USA}
\affiliation{SLAC National Accelerator Laboratory, Menlo Park, CA 94025, USA}
\author{R.~A.~Gruendl}
\affiliation{Department of Astronomy, University of Illinois at Urbana-Champaign, 1002 W. Green Street, Urbana, IL 61801, USA}
\affiliation{National Center for Supercomputing Applications, 1205 West Clark St., Urbana, IL 61801, USA}
\author{J.~Gschwend}
\affiliation{Laborat\'orio Interinstitucional de e-Astronomia - LIneA, Rua Gal. Jos\'e Cristino 77, Rio de Janeiro, RJ - 20921-400, Brazil}
\affiliation{Observat\'orio Nacional, Rua Gal. Jos\'e Cristino 77, Rio de Janeiro, RJ - 20921-400, Brazil}
\author{G.~Gutierrez}
\affiliation{Fermi National Accelerator Laboratory, P. O. Box 500, Batavia, IL 60510, USA}
\author{K.~Honscheid}
\affiliation{Center for Cosmology and Astro-Particle Physics, The Ohio State University, Columbus, OH 43210, USA}
\affiliation{Department of Physics, The Ohio State University, Columbus, OH 43210, USA}
\author{K.~Kuehn}
\affiliation{Australian Astronomical Optics, Macquarie University, North Ryde, NSW 2113, Australia}
\affiliation{Lowell Observatory, 1400 Mars Hill Rd, Flagstaff, AZ 86001, USA}
\author{N.~Kuropatkin}
\affiliation{Fermi National Accelerator Laboratory, P. O. Box 500, Batavia, IL 60510, USA}
\author{M.~A.~G.~Maia}
\affiliation{Laborat\'orio Interinstitucional de e-Astronomia - LIneA, Rua Gal. Jos\'e Cristino 77, Rio de Janeiro, RJ - 20921-400, Brazil}
\affiliation{Observat\'orio Nacional, Rua Gal. Jos\'e Cristino 77, Rio de Janeiro, RJ - 20921-400, Brazil}
\author{J.~L.~Marshall}
\affiliation{George P. and Cynthia Woods Mitchell Institute for Fundamental Physics and Astronomy, and Department of Physics and Astronomy, Texas A\&M University, College Station, TX 77843,  USA}
\author{F.~Menanteau}
\affiliation{Department of Astronomy, University of Illinois at Urbana-Champaign, 1002 W. Green Street, Urbana, IL 61801, USA}
\affiliation{National Center for Supercomputing Applications, 1205 West Clark St., Urbana, IL 61801, USA}
\author{R.~Miquel}
\affiliation{Instituci\'o Catalana de Recerca i Estudis Avan\c{c}ats, E-08010 Barcelona, Spain}
\affiliation{Institut de F\'{\i}sica d'Altes Energies (IFAE), The Barcelona Institute of Science and Technology, Campus UAB, 08193 Bellaterra (Barcelona) Spain}
\author{A.~Palmese}
\affiliation{Fermi National Accelerator Laboratory, P. O. Box 500, Batavia, IL 60510, USA}
\affiliation{Kavli Institute for Cosmological Physics, University of Chicago, Chicago, IL 60637, USA}
\author{F.~Paz-Chinch\'{o}n}
\affiliation{Institute of Astronomy, University of Cambridge, Madingley Road, Cambridge CB3 0HA, UK}
\affiliation{National Center for Supercomputing Applications, 1205 West Clark St., Urbana, IL 61801, USA}
\author{A.~A.~Plazas}
\affiliation{Department of Astrophysical Sciences, Princeton University, Peyton Hall, Princeton, NJ 08544, USA}
\author{A.~Roodman}
\affiliation{Kavli Institute for Particle Astrophysics \& Cosmology, P. O. Box 2450, Stanford University, Stanford, CA 94305, USA}
\affiliation{SLAC National Accelerator Laboratory, Menlo Park, CA 94025, USA}
\author{E.~Sanchez}
\affiliation{Centro de Investigaciones Energ\'eticas, Medioambientales y Tecnol\'ogicas (CIEMAT), Madrid, Spain}
\author{V.~Scarpine}
\affiliation{Fermi National Accelerator Laboratory, P. O. Box 500, Batavia, IL 60510, USA}
\author{M.~Schubnell}
\affiliation{Department of Physics, University of Michigan, Ann Arbor, MI 48109, USA}
\author{S.~Serrano}
\affiliation{Institut d'Estudis Espacials de Catalunya (IEEC), 08034 Barcelona, Spain}
\affiliation{Institute of Space Sciences (ICE, CSIC),  Campus UAB, Carrer de Can Magrans, s/n,  08193 Barcelona, Spain}
\author{I.~Sevilla-Noarbe}
\affiliation{Centro de Investigaciones Energ\'eticas, Medioambientales y Tecnol\'ogicas (CIEMAT), Madrid, Spain}
\author{M.~Smith}
\affiliation{School of Physics and Astronomy, University of Southampton,  Southampton, SO17 1BJ, UK}
\author{M.~Soares-Santos}
\affiliation{Department of Physics, University of Michigan, Ann Arbor, MI 48109, USA}
\author{E.~Suchyta}
\affiliation{Computer Science and Mathematics Division, Oak Ridge National Laboratory, Oak Ridge, TN 37831}
\author{M.~E.~C.~Swanson}
\affiliation{National Center for Supercomputing Applications, 1205 West Clark St., Urbana, IL 61801, USA}
\author{G.~Tarle}
\affiliation{Department of Physics, University of Michigan, Ann Arbor, MI 48109, USA}
\author{J.~Weller}
\affiliation{Max Planck Institute for Extraterrestrial Physics, Giessenbachstrasse, 85748 Garching, Germany}
\affiliation{Universit\"ats-Sternwarte, Fakult\"at f\"ur Physik, Ludwig-Maximilians Universit\"at M\"unchen, Scheinerstr. 1, 81679 M\"unchen, Germany}

\collaboration{DES Collaboration}

% \begin{abstract}
% This is a sample document created by \texttt{mkauthlist v1.2.4}.
% \end{abstract}
% \maketitle
% \end{document}

\preprint{DES-2019-0458}
\preprint{FERMILAB-PUB-20-421-AE}

\label{firstpage}

\begin{abstract}
We describe perturbation theory (PT) models of galaxy bias for applications to photometric galaxy surveys. We model the galaxy-galaxy and galaxy-matter correlation functions in configuration space and validate against measurements from mock catalogs designed for the Dark Energy Survey (DES). We find that an effective PT model with five galaxy bias parameters provides a good description of the 3D correlation functions above scales of 4 Mpc/$h$ and $z < 1$. Our tests show that at the projected precision of the DES-Year 3 analysis, two of the non-linear bias parameters can be fixed to their co-evolution values, and a third (the $k^2$ term for higher derivative bias) set to zero. The agreement is typically at the 2 percent level over scales of interest, which is the statistical uncertainty of our simulation measurements. To achieve this level of agreement, our {\it fiducial} model requires using the full non-linear matter power spectrum (rather than the 1-loop PT one). We also measure the relationship between the non-linear and linear bias parameters and compare them to their expected co-evolution values. We use these tests to motivate the galaxy bias model and scale cuts for the cosmological analysis of the Dark Energy Survey;  our conclusions are generally applicable to all photometric surveys.
\end{abstract}

\maketitle
\section{Introduction}
\label{sec:intro}
The structure in the universe at low redshift was seeded by small perturbations in the early universe. Although the evolution of these tiny perturbations is well described in the linear regime, their non-linear evolution on small scales is an active area of research. 

There is a well-formulated framework of non-linear perturbative expansions of these early fluctuations in both Eulerian and Lagrangian space (see \cite{Bernardeau:2001qr} and \cite{Desjacques_2018} for a review).  Major approaches include Standard Perturbation Theory (SPT,  \cite{Jain94, Goroff86}), Lagrangian Perturbation Theory (LPT, \cite{Bouchet95, Matsubara08}), Renormalized Perturbation Theory (\cite{Crocce06}), Effective Field Theory (EFT, \cite{Carrasco:2012cv, Vlah15, perko2016biased}). Although these theories analytically describe the relation between dark matter non-linear density perturbations and linear density perturbations, direct observations exist only for some biased tracers of the underlying dark matter field. These theories have therefore been extended to describe biased tracers like galaxies \citep{Fry_93, Coles93, Heavens_98, Scherrer_98, Matsubara08, McDonald2009, Matsubara2013, Carlson13} and applied to data \citep{Blake_2011,Mar_n_2013,Sanchez_2016,Gil_Marin_2016,Beutler_2016,Grieb_2017, DAmico_eft:2019, Ivanov_eft:2019}.

Another analytical approach for biased tracers is the halo model framework (see \citep{CooraySheth02} for a review). The halo model assumes that all matter is bound in virialized objects (halos) and relates clustering statistics to halos. This framework can be extended to include the observed tracers, for example, via the Halo Occupation Distribution (HOD) (\citep{BerlindWeinberg02, Zheng05}).  However, unlike the perturbation theory, the parameterization of the HOD is tracer dependent and cannot be easily generalized. Moreover, the HOD only describes the distribution of galaxies inside halos (known as the 1-halo term). To correctly describe the clustering of galaxies on weakly non-linear scales, between the non-linear 1-halo regime and the large scale linear regime, would require a combination with perturbative models.

Several studies have tested the perturbation theory (PT) of biased tracers in Fourier space (mostly focused on redshift surveys) \citep{Saito2014a, Angulo_2015, bella2018impact, Werner_2019, alex2020testing}. This study focuses on PT in configuration space using Standard Perturbation Theory (SPT) and Effective Field Theory (EFT). We use the 3D correlation functions, $\xi_{\mathrm{gg}}$ and $\xi_{\mathrm{gm}}$, constructed from galaxy and matter catalogs built from  simulations. One of the key results of our analysis is the minimum length scale for which the correlation functions can be modeled with PT.

The mock catalogs used in this analysis are designed for the Dark Energy Survey (DES). As described in Section \ref{sec:data}, our focus is on Year 3 (Y3) DES data sets, for which we use the mocks to validate our PT models. This data set constitutes the largest current imaging survey of galaxies, and thus careful testing and validation that matches its statistical power are essential for extracting information in the non-linear regime. We also project the 3D correlations from mocks to the angular correlations (as measured by photometric surveys), but since projection results in loss of information, our 3D tests are more stringent. Since the PT formalism is not tied to any particular tracer, and the scales of interest are well above the 1-halo regime (where differences in galaxy assignment schemes matter), we expect that our conclusions will have broad validity for the lensing and galaxy clustering analyses from imaging surveys.

We also aim to test the accuracy of different variants of perturbation theory for cosmological applications with DES. Although this analysis is at fixed cosmology, we implement fast evaluations of the projected correlations so that they can feasibly be used for cosmological parameter analysis. Finally, we explore the possibility of placing well-motivated priors on some of the PT bias parameters.  

This paper is organized as follows. In Sec.~\ref{sec:formalism}, we review the existing perturbation theory literature and the models used in this study. Sec.~\ref{sec:data}
describes the simulations used for the measurements and Sec.~\ref{sec:analysis} the analysis choices. The results are presented in Sec.~\ref{sec:results}, and we conclude in Sec.~\ref{sec:conclusion}.

\section{Formalism}
\label{sec:formalism}
We summarize in this section the perturbation theory formalism used in our study and the projected two-point statistics relevant for surveys like DES. We are interested in modeling both the matter and galaxy distribution. Different perturbation theory approaches describe the evolved galaxy density fluctuations $ \delta_{\mathrm{g}}(\textbf{x}) $ of a biased tracer, $ g $, in terms of the linear matter density fluctuations $ \delta_{\mathrm{L}}(\textbf{x})$. Although formally the relationship between $\delta_{\mathrm{g}}(\textbf{x})$ and $\delta_{\mathrm{L}}(\textbf{x})$ is on the full past Lagrangian path of a particle at Eulerian position $\textbf{x}$, in this analysis we use the approximation that this relationship is instantaneous, meaning $\delta_{\mathrm{g}}(\textbf{x},z)$ is related only to $\delta_{\mathrm{L}}(\textbf{x},z)$ at any redshift $z$.

\subsection{Standard Perturbation Theory}\label{sec:spt}
Standard perturbation theory expands the evolved dark matter density field, $\delta_{\mathrm{m}}(\textbf{x})$ in terms of the extrapolated linear density field, shear field, the divergence of the velocity field and rotational invariants constructed using the gravitational potential. In Fourier space, this expansion can be written as \cite{Bernardeau:2001qr}

\begin{multline}
\delta_{\mathrm{m}}(\bk) = \sum \frac{1}{n!} \int \frac{d^3 k_1}{(2\pi)^3} ... \frac{d^3 k_n}{(2\pi)^3} (2\pi)^3 \delta_{\mathrm{D}} (\bk_{1..n} - \bk) \\ F_n(\bk_1,..,\bk_n) \delta_{\mathrm{L}}(\bk_1)...\delta_{\mathrm{L}}(\bk_n)\,.
\end{multline}\label{eq:deltam_F2}

Here $ F_n(\bk_1,..,\bk_n) $ are the mode coupling kernels constructed out of correlations between the scalar quantities mentioned above and $\delta_{\mathrm{D}}$ is the Dirac delta function. The form of the $ F_n $ kernels can be derived by solving the perturbative fluid equations.
For example under the assumptions of the spatially flat, cold dark matter model of cosmology, $ F_2 $ is well approximated by
\begin{equation}\label{key}
F_2(\textbf{k},\textbf{k}') = \bigg[(1 + \alpha) + \mu \bigg(\frac{k}{k'} + \frac{k'}{k}\bigg) + (1 - \alpha) \mu^2\bigg]\,.
\end{equation}
For $\Omega_{\mathrm{m}} < 1 $, $\alpha = \frac{3}{7}(\Omega_{\mathrm{m}})^{-2/63} $ and $\mu = \frac{\textbf{k} \cdot \textbf{k}'}{k \cdot k'} $. In this analysis, we use the Einstein de-Sitter limit and assume $\alpha = \frac{3}{7}$.

\subsubsection{Biased tracers}

The overdensity of biased tracers is modeled as the sum of a deterministic function of the dark matter density ($f\left[\delta_m(\textbf{x})\right]$) and a stochastic component ($\varepsilon(\textbf{x})$)
\begin{equation}\label{eq:delg_gen}
\delta_{\mathrm{g}}(\textbf{x}) = f\left[\delta_m(\textbf{x})\right] + \varepsilon(\textbf{x})\,.
\end{equation}

In this analysis we ignore the stochastic contribution and focus on the deterministic relation between the dark matter field and the biased tracer. Assuming a local biasing scheme, this expansion is given as (\cite{Goroff86}) 
\begin{equation}
\delta^{\rm local}_{\mathrm{g}}(\textbf{x}) = \sum_{n=1}^{\infty} \frac{b_n}{n!} \delta_{\mathrm{m}}^n(\textbf{x})\,.
\end{equation}

However, as is well known (\citep{Fry_93, Scherrer_98}), on small scales this local biasing in Eulerian space rapidly breaks down. Assuming isotropy and homogeneity, the bias parameters have to be scalar and hence the density of a tracer can only depend on scalar quantities (\citep{McDonald2009}). Therefore, non-local terms can only be sourced by scalar quantities constructed out of gravitational evolution of matter density ($\delta_{\mathrm{m}}$), shear ($\nabla_i \nabla_j \Phi$) and velocity divergences ($\nabla_i v_j$). Following the procedure in \citep{McDonald2009, Eggemeier_2019,Chan_2012}, these contributions can be re-arranged into independent terms that contribute to the overdensity of galaxies ($\delta_g$) at different orders  
\begin{multline}\label{eq:delg_spt3}
    \delta_{\mathrm{g}} \sim f(\delta_{\mathrm{m}}, \nabla_i \nabla_j \Phi, \nabla_i v_j) \sim f^{(1)}(\delta_{\mathrm{m}}) + f^{(2)}(\delta^2_{\mathrm{m}}, s^2)  \\ + f^{(3)}(\delta^3_{\mathrm{m}},\delta_{\mathrm{m}} s^2, \psi, st) + ...\,.
\end{multline}

Here $f^{i}$ are the functions that contribute to the total overdensity at $i$-th order only and $\psi,s$ and $t$ are the scalar quantities constructed out of shear and velocity divergences. When expanding the form of these function $f^{i}$ up to third order, we introduce un-normalized bias factors as given in Eq 9 and Eq 12 of \citet{McDonald2009}. In Fourier space, the equivalent equation is Eq.~(A14) of \citet{Saito2014a}.

\subsection{Higher derivative bias}

In the above section, the non-local terms included in the expansion of galaxy overdensity comes only from shear and velocity divergences. However, those terms are still local in the spatial sense, meaning that the formation of biased tracers only depends on the scalar quantities discussed above at the same position as the tracer. A short-range non-locality due to non-linear effects in  halo and galaxy formation within some some scale $R$, will change  Eq.~\ref{eq:delg_gen} to: (\citealp{McDonald2009})
\begin{equation}\label{eq:delg_eft}
\delta_{\mathrm{g}}(\textbf{x}) = f\left[\delta_{\mathrm{m}}(\textbf{x}')\right]\,,
\end{equation}
where, generally $|\textbf{x} - \textbf{x}'| < R$ and $R$ is usually of the order of halo radius. Taylor expanding this function we can see that lowest order gradient-type term that can contribute to $\delta_{\mathrm{g}}$ is proportional to $\nabla^2 \delta_{\mathrm{m}}$. Hence, we can further generalize our Eq.~\ref{eq:delg_spt3} to include this gradient-type term as
\begin{multline}\label{eq:delg_full}
    \delta_{\mathrm{g}} \sim f(\delta_{\mathrm{m}}, \nabla_i \nabla_j \Phi, \nabla_i v_j) \sim f^{(1)}(\delta_{\mathrm{m}}) + f^{(2)}(\delta^2_{\mathrm{m}}, s^2)  \\ + f^{(3)}(\delta^3_{\mathrm{m}},\delta_{\mathrm{m}} s^2, \psi, st) + f^{\rm grad}(\nabla^2 \delta_{\mathrm{m}}) + ...\,.
\end{multline}

Note that in Fourier space, this term would scale as $k^2 \delta_{\mathrm{m}} (k)$.

\subsection{Effective Field Theory}\label{sec:eft}
Moreover, as discussed in \citet{Carrasco:2012cv}, it is theoretically inconsistent to use small scale modes in the integration over Fourier space. So we use effective integrated ultra-violet (UV) terms in the final expansion for the power spectrum. This effective term also enters as a $ k^2 $ contribution in the large-scale limit. For example, if we expand the non-linear matter power spectrum in terms of the linear power spectrum ($P_{\rm L}(k)$) using the PT framework, we have to include this $k^2$ piece usually written as $c^2_{\rm s} k^2 P_{\rm L}(k)$, where $c_{\rm s}$ is the effective adiabatic sound speed. 

\subsection{Regularized PT power spectra}\label{sec:regPT}

Note that the bias parameters that will appear in the expansion of $\delta_g$ in Eq.~\ref{eq:delg_full} will be un-observable ``bare bias" parameters and need not have the physical meaning usually attributed to the large scale tracer bias (for example, \ the measurable responses of galaxy statistics to a given fluctuation). We refer the reader to \citet{McDonald2009} for the details on the renormalization of these ``bare bias" parameters by combining all the parameters with similar power spectrum kernels. After renormalizing, we can write the tracer-matter cross spectrum ($P_{\mathrm{gm}}$) and auto power spectrum of the tracer ($P_{\mathrm{gg}}$) as:

\begin{multline}\label{eq:P_tm}
P_{\mathrm{gm}}(k)= b_1 P_{\mathrm{mm}}(k) +  \frac{1}{2} b_2P_{\rm b_1 b_2}(k) + \frac{1}{2} b_{\mathrm{s}}P_{\rm b_1 s^2}(k) + \\ \frac{1}{2} b_{\rm 3nl}P_{\rm b_1 b_{\rm 3nl} }(k) + (b^{\rm hd}_{\nabla^2 \delta} + c^2_{\rm s}) k^2 P^{\rm grad}_{\mathrm{mm}}(k)\,.
\end{multline}

\begin{multline}
	P_{\mathrm{gg}}(k) = b_1^2 P_{\mathrm{mm}}(k) + b_1b_2 P_{\rm b_1 b_2}(k) + b_1b_{\mathrm{s}}P_{\rm b_1 s^2}(k) + \\ b_1b_{\rm 3nl}P_{\rm b_1 b_{\rm 3nl} }(k) +  \frac{1}{4}b_2^2 P_{\rm b_2 b_2}(k) + 
	\frac{1}{2}b_2b_{\mathrm{s}}P_{\rm b_2 s^2}(k) + \\ \frac{1}{4}b_{\mathrm{s}}^2 P_{\rm s^2 s^2}(k)  + b_1 (2 b^{\rm hd}_{\nabla^2 \delta} + c^2_{\rm s})  k^2 P^{\rm grad}_{\mathrm{mm}}(k)\,. 
\end{multline}

Here the bias parameters like $ b_1 $, $ b_2 $,	$ b_{\mathrm{s}} $ and $ b_{\rm 3nl} $ are the renormalized bias parameters which are physically observable. 
The bias parameter $b^{\rm hd}_{\nabla^2 \delta}$ is the higher-derivative bias parameter and $c^2_{\rm s}$ is the sound speed term as described by EFT (\S\ref{sec:eft}). As for the kernels, $P_{\rm b_1 b_2}(k)$ is generated from ensemble average of $\langle \delta_{\mathrm{m}} \delta^2_{\mathrm{m}} \rangle$, $P_{\rm b_1 s^2}(k)$ is generated from $\langle \delta_{\mathrm{m}} s^2 \rangle$ and $P_{\rm b_1 b_{\rm 3nl} }$ is generated from a combination of ensemble average between $\delta_{\mathrm{m}}$ and arguments of $f^{(3)}$ (see Eq.~\ref{eq:delg_full}) that contribute at 1-loop level \citep{Saito2014a}. 
For the exact form of above kernels, see the Appendix A  of \citet{Saito2014a}. 

Instead of expanding the Eulerian galaxy overdensity field directly as we have done above, we can also predict the galaxy overdensity by evolving the Lagrangian galaxy overdensity (see \citet{Matsubara2013} for detailed calculations). These two approaches should evaluate to the same galaxy overdensity at a given loop order \citep{Fry_1996,Baldauf_2012,Chan_2012,Matsubara2013,Saito2014a}. By equating the two approaches and neglecting shear-like terms in the Lagrangian overdensity as they are small for bias values of our interest (see \S\ref{sec:results} and \cite{Modi_2017}), we get the prediction of the co-evolution value of the renormalized bias parameters: $b_{\mathrm{s}} = (-4/7) \times (b_1 - 1)$ and $b_{\rm 3nl} = (b_1 - 1)$\footnote{note that our co-evolution value of  $b_{\rm 3nl}$ differs from \citet{Saito2014a} as we include their prefactor of 32/315 in our definition of $P_{\rm b1b3nl}$} \citep{Matsubara2013, Saito2014a}. This co-evolution picture naturally describes how gravitational evolution generates the non-local biasing even from the local biased tracers in high redshift Lagrangian frame.

We use different choices of $P_{\mathrm{mm}}$ and $P^{\rm grad}_{\mathrm{mm}}$ in our analysis. These choices will be detailed in the \S\ref{sec:dv_models}.

\subsection{3D statistics to projected statistics}\label{sec:3d_to_2d}

We are interested in the cosmological applications of imaging surveys via projected correlation functions. Projections of the 3D correlation functions $\xi_{\mathrm{gg}}$ and $\xi_{\mathrm{gm}}$, to angular coordinates in finite redshift bins give the projected correlations known as $w_{\mathrm{gg}}(\theta)$ and $\gamma_t(\theta)$ respectively. We estimate the covariance of these projected statistics for the DES-Y3 like survey. This allows us to estimate the angular scales for which our perturbation theory model is a good description for DES-Y3 like sensitivity.

\subsubsection{Galaxy-Galaxy clustering}

The angular correlation function $w_{\mathrm{gg}}(\theta)$ is given by the Limber integral
\beq\label{eq:wgg_t}
w_{\mathrm{gg}}(\theta) = \int_0^{\infty} d\chi \ \chi^4 \ \phi^2(\chi) \int_{-\infty}^{\infty} dr_{\parallel} \ 
\xi_{\mathrm{gg}} \bigg(\sqrt{r_{\parallel}^2 + \chi^2 \theta^2} \bigg),
\eeq
where $\chi$ is the comoving distance and $\phi(\chi)$ is the normalized radial selection function of the lens galaxies,  related to the normalized redshift distribution of lens galaxies ($n_{\mathrm{g}}(z)$)  as $\phi(\chi) = (1/\chi^2)(dz/d\chi) n_{\mathrm{g}}(z)$.

To simplify the above equation and ones that follow, the inner integral will be denoted by $w_{\mathrm{gg}}^p = \int_{-\infty}^{\infty} dr_{\parallel} \ 
\xi_{\mathrm{gg}} \bigg(\sqrt{r_{\parallel}^2 + \chi^2 \theta^2} \bigg)$. A similar equation applies for the galaxy-matter correlation as well.  The integral limits for this projection integral are from $-\infty$ to $\infty$. Though our analysis of survey data is over a finite projection length, as described below in \S\ref{sec:data}, our thinnest tomographic bin spans redshift  $0.3 < z < 0.45$ -- a distance of over 500 Mpc/$h$. Moreover, as our analysis uses true galaxy redshifts, there is no peculiar velocity effect on projected integrals \citep{Bosch_2013}. Therefore ignoring the finite bin size introduces negligible errors in our correlation function predictions.

Substituting the radial selection function in terms of the  galaxy redshift distribution and using the above definition of $w^p$, the projected galaxy clustering, $w_{\mathrm{gg}}(\theta)$, can be expressed as
\beq\label{eq:wgg_t_simp}
w_{\mathrm{gg}}(\theta) = \int_0^{\infty} dz \ \frac{dz}{d\chi} \  n^2_{\mathrm{g}} (z) \ w_{\mathrm{gg}}^p (\chi \theta)\,.
\eeq

\subsubsection{Galaxy-galaxy lensing}

The galaxy-galaxy lensing signal ($\gamma_{\rm t}$) is related to the excess surface mass density ($\Delta \Sigma$) around lens galaxies by

\beq
\gamma_{\rm t}(\theta, z_{\rm l}, z_{\rm s})  = \frac{\Delta \Sigma (\theta, z_{\rm l})}{\Sigma_{\rm crit} (z_{\rm l}, z_{\rm s})}\,, 
\eeq
where $\Sigma_{\rm crit}$ is the critical surface mass density given by
\beq
\Sigma_{\rm crit} (z_{\rm l}, z_{\rm s}) = \frac{c^2}{4\pi G} \frac{D_{\mathrm{A}}(z_{\mathrm{s}})}{D_{\mathrm{A}}(z_{\rm l})D_{\mathrm{A}}(z_{\mathrm{l}},z_{\mathrm{s}})}\,.
\eeq
Here $D_{\mathrm{A}}$ is the angular diameter distance, $z_{\mathrm{l}}$ is the redshift of the lens and $z_{\mathrm{s}}$ is the redshift of the source. 

The surface mass density at the projected distance $r_{\rm p} = \chi \theta$ can be related to the projected galaxy-matter correlation function by
\beq
\Sigma(r_{\rm p},z) = \langle \Sigma \rangle + \rho_{\mathrm{m}}(z)\  w_{\mathrm{gm}}^p(r_{{p}},z)\,,
\eeq
where $\langle \Sigma \rangle$ is the mean surface density
\beq
 \langle \Sigma \rangle = \int_{z_{\rm min}}^{z_{\rm max}} dz \ \frac{d\chi}{dz}  \ \rho_{\mathrm{m}}(z)\,,
\eeq
and $\rho_{\mathrm{m}}(z) = \Omega_{{\mathrm{m}},0} (1+z)^3 \rho_{\rm crit,0}$ is the mean density of the universe.

Therefore, the excess surface density is
\begin{align}
\Delta \Sigma(r_{\rm p},z) &= \rho_{\mathrm{m}}(z) (\bar{w}_{\mathrm{gm}}^p(r_{\rm p},z) - w_{\mathrm{gm}}^p(r_{\rm p},z)) \\ 
& = \rho_{\mathrm{m}}(z) \Delta w_{\mathrm{gm}}^p(r_{\rm p},z)\,
\end{align}
where, $\bar{w}_{\mathrm{gm}}^p(r_{\rm p},z)$ is given as:
\beq\label{eq:wp_bar_gm}
\bar{w}_{\mathrm{gm}}^p (\chi \theta, z) = \frac{2}{(\chi \theta)^2}\bigg[\int_0^{\chi \theta} dr_p \ r_p \ \ w_{\mathrm{gm}}^p (r_p, z) \bigg]\,.
\eeq

Now combining all the above equations, the galaxy-galaxy lensing signal for lenses at redshift $z_{\mathrm{l}}$ and sources at redshift $z_{\mathrm{s}}$ is
\beq
\gamma_{\rm t} (\theta, z_{\rm l}, z_{\rm s}) = \frac{\Delta w_{\mathrm{gm}}^p (\chi \theta, z_{\rm l}) \ \rho_{\mathrm{m}}(z_{\rm l})}{\Sigma_{\rm crit}(z_{\rm l}, z_{\rm s})}\,.
\eeq

Averaging this signal with the redshift distribution of sources ($n_{\mathrm{s}}(z_{\mathrm{s}})$) would give
\beq
\gamma_{\rm t} (\theta, z_{\mathrm{l}}) = \Delta w_{\mathrm{gm}}^p (\chi \theta) \ \rho_{\mathrm{m}}(z) \int_{0}^{\infty} dz_{\mathrm{s}} \ n_{\mathrm{s}}(z_{\mathrm{s}}) \ \frac{1}{\Sigma_{\rm crit}(z_{\mathrm{l}}, z_{\mathrm{s}})}\,.
\eeq

Finally, averaging this signal with the redshift distribution of lens galaxies ($n_{\mathrm{g}}(z_{\mathrm{l}})$) gives
\begin{multline}\label{eq:gt}
   \gamma_{\rm t} (\theta) =   \int_{0}^{\infty} dz_{\mathrm{l}} \ \rho_{\mathrm{m}}(z_{\mathrm{l}}) \ n_{\mathrm{g}}(z_{\mathrm{l}}) \ \Delta w_{\mathrm{gm}}^p (\chi \theta) \\ \times \int_{0}^{\infty} dz_{\mathrm{s}} \ n_{\mathrm{s}}(z_{\mathrm{s}}) \ \frac{1}{\Sigma_{\rm crit}(z_{\mathrm{l}}, z_{\mathrm{s}})}\,.
\end{multline}

The tangential shear $\gamma_{\rm t}(\theta)$ is nonlocal and depends on the correlation function at all scales smaller than the transverse distance $\chi \theta$ (Eq.~\ref{eq:wp_bar_gm}, see \citet{MacCrann:2019ntb, Baldauf_2010} for a detailed analysis). Perturbation theory is not adequate for modeling these small scales. We therefore add to $\gamma_{\rm t}$ a term representing a point mass contribution: $B/\theta^2$, where $B$ is the average point-mass for a sample of lens and source galaxies and is treated as a free parameter. Any spherically symmetric mass distribution within the minimum scale used is captured by the point mass term, thus removing our sensitivity to these scales. Our final expression for the galaxy-galaxy lensing signal is
\begin{equation}\label{eq:gt_b}
    \gamma_{\rm t} (\theta) =  \gamma^{\rm theory}_{\rm t} (\theta) + \frac{B}{\theta^2}\,,
\end{equation}
with $\gamma^{\rm theory}_{\rm t}$ given by Eq.~\ref{eq:gt}.

\section{Simulations and mock catalogs}
\label{sec:data}

The full DES survey was completed in 2019 and covered $\sim5000$ square degrees of the South Galactic Cap. Mounted on the Cerro Tololo Inter-American Observatory (CTIO) $4~m$ Blanco telescope in Chile, the 570-megapixel Dark Energy Camera \citep[DECam][]{Flaugher15} images the field in $grizY$ filters. The raw images are processed by the DES Data Management (DESDM) team  \citep{Sevilla11,Morganson18}. The Year 3 (Y3) catalogs of interest for this study span the full footprint of the survey but with fewer exposures than the complete survey. About 100 million galaxies have shear and photometric redshift measurements that enable their use for cosmology. For the full details of the data and the galaxy and lensing shear catalogs, we refer the readers to \cite{Sevilla:inprep} and \cite{Sheldon:inprep}.

We use DES-like mock galaxy catalogs from the \mice simulation suite in this analysis. The \mice  Grand Challenge simulation (MICE-GC) is an N-body simulation run in a cube with side-length 3 Gpc/$h$ with $4096^3$ particles  using the Gadget-2 code \citep{Gadget2} with mass resolution of $2.93 \times 10^{10} M_{\odot}/h$. Halos are identified using a Friend-of-Friends algorithm with linking length 0.2. For further details about this simulation, see \citet{Fosalba:2015a}. These halos are then populated with galaxies using a hybrid sub-halo abundance matching plus halo occupation distribution (HOD) approach, as detailed in \citet{MICE:2}. These methods are designed to match the joint distributions of luminosity, $g-r$ color, and clustering amplitude observed in SDSS \citep{SDSS:Zehavi_2011}. The construction of the halo and galaxy catalogs is described in \citet{Crocce:2015}. \mice assumes a flat $\Lambda$CDM cosmological model with  $h=0.7$, $\Omega_{\rm m}=0.25$, $\Omega_{\mathrm{b}}=0.044$ and $\sigma_8=0.8$. 

We use two galaxy samples generated from the full \mice galaxy catalog. A DES-like lightcone catalog of \redmagic galaxies \citep{Rozo_2016} with average photometric errors matching DES Y1 data is generated. We also use another galaxy sample (\maglim hereafter) based on cuts on galaxy magnitude only. This sample is created by imposing a cut on the simulated DES i-band like magnitudes (mag-i) of \mice galaxies \citep{Porredon:inprep}. The galaxies in this \maglim \ sample follow the conditions: mag-i$ > 17.5$ and mag-i$ < 4z + 18$ where $z$ is the true redshift of the galaxy. This definition results from a sample optimization process when deriving cosmological information from a combined clustering and lensing analysis \citep{Porredon:inprep}. Both simulated galaxy samples populate one octant of the sky (ca. 5156 sq. degrees), which is slightly larger than the sky area of DES Y3 data (approximately 4500 sq. degrees, \cite{Sheldon:inprep}). From these simulations, we measure the non-linear bias parameters at fixed cosmology, which we use as fiducial values for the DES galaxy sample(s).

\begin{figure}
    \centering
    \includegraphics[width=1.0\linewidth]{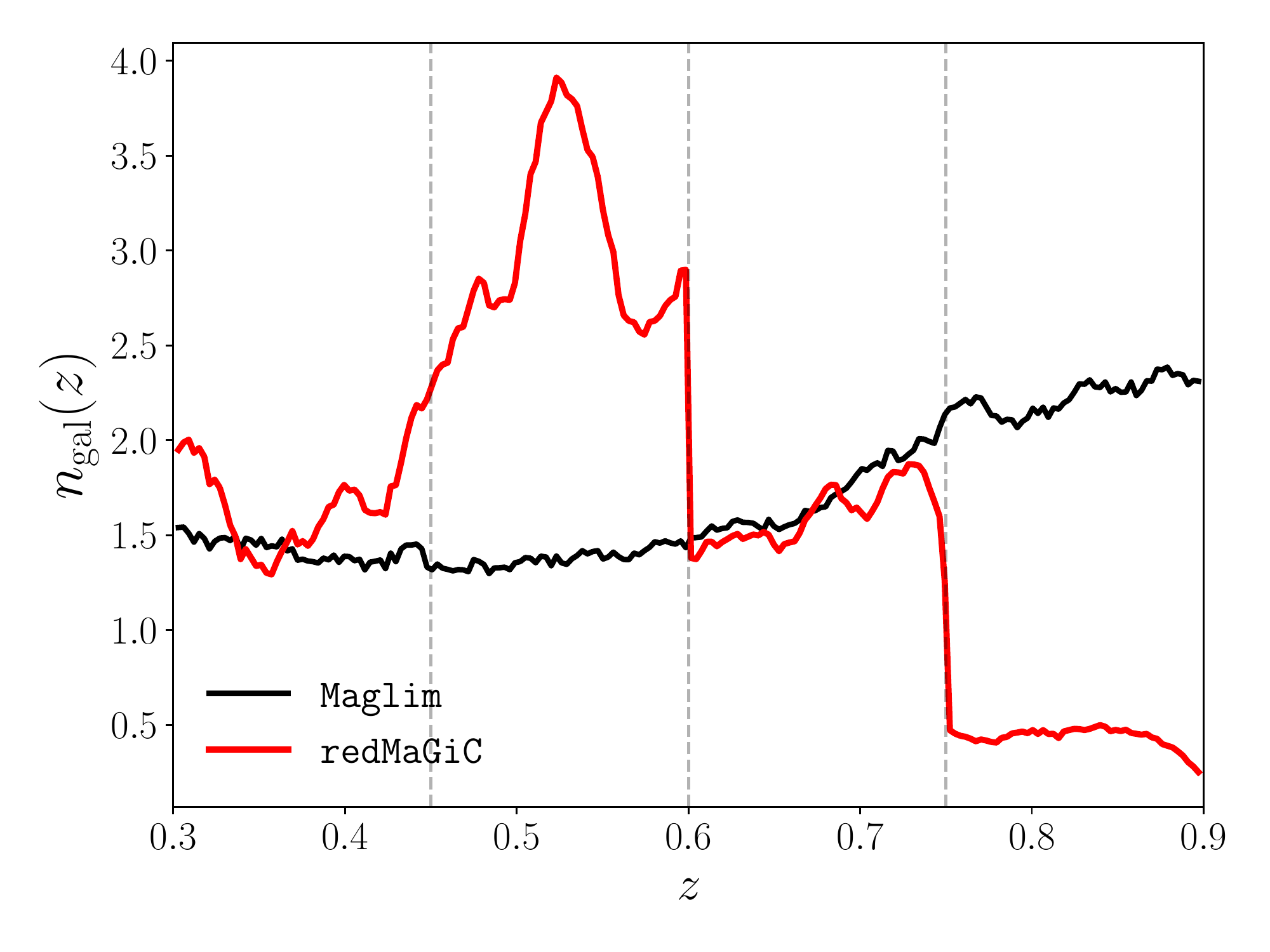}
    \caption{Comparison of normalized number density of galaxies corresponding to \redmagic and \maglim samples. The dashed vertical lines denote the tomographic bin edges. }
    \label{fig:nzcomp}
\end{figure}

As detailed in later sections, we divide our galaxy samples into four tomographic bins with edges $[0.3,0.45,0.6,0.75,0.9]$. These bins are the same as the last four of the five tomographic bins used in the DES Y1 analysis \citep{DESy1,DES:Bias}. We do not fit to the first tomographic bin of DES Y1 analysis (which is $0.15 < z < 0.3$) because we are limited by the jackknife covariance estimate (see \S\ref{sec:covaraince} and Appendix \ref{app:cov}). These tomographic bins cover a similar redshift range as planned for the DES Y3 analysis.  Note that we bin our galaxies used in this analysis using their true spectroscopic redshift. Therefore there is no overlap in the redshift distribution of galaxies between two different bins. After all color, magnitude, and redshift cuts, there are 2.1 million \redmagic galaxies and 2.0 million \maglim galaxies (downsampled to have approximately the same number density as \redmagic)  used in this analysis. The normalized number densities of two catalogs are shown in Fig.~\ref{fig:nzcomp}.

We note that although both the mock catalogs used in this analysis are calibrated with DES Y1 data, we do not expect our tests and conclusions to change with Y3 mock catalog. Since our tests are based on the true redshifts of the galaxies, we are not sensitive to photometric redshift uncertainties, exact tomography choices, and color selection of the galaxies.

\section{Analysis}
\label{sec:analysis}

\subsection{Data Vector and Models}\label{sec:dv_models}

Our main analysis involves the auto and cross-correlations functions for galaxies and matter: $\xi_{\mathrm{mm}}$, $\xi_{\mathrm{gm}}$ and $\xi_{\mathrm{gg}}$. Our focus is on galaxy bias, so we would like to minimize artifacts that are specific to the clustering of matter, in particular sampling effects due to the finite volume of the simulations (see Appendix~\ref{app:cov}).  Therefore, we fit our theory models to the ratios:  $\xi_{\mathrm{gg}}/\xi_{\mathrm{mm}}$ and $\xi_{\mathrm{gm}}/\xi_{\mathrm{mm}}$ so that the galaxy two-point functions are analyzed relative to the matter-matter correlation (see Appendix~\ref{app:fit_corr} and Fig.~\ref{fig:gg_mm_comp_bin3} for an analysis on correlation functions $\xi_{\mathrm{gm}}$ and $\xi_{\mathrm{gg}}$ directly). 
% The measurements of these ratios are shown in Fig.~\ref{fig:xi_gg_gm}. 
We consider three models to describe these measured ratios:

\begin{equation}\label{eq:modelABC}
\left.\begin{aligned}
 {\rm A} : \frac{\xi_{\mathrm{gm}}}{\xi_{\mathrm{mm}}}&= b_1\\
 {\rm B} : \frac{\xi_{\mathrm{gm}}}{\xi_{\mathrm{mm}}}&= \frac{\mathcal{F}\bigg[b_1 P^{\rm 1-loop}_{\mathrm{mm}}(k) + P^{\rm 1-Loop}_{\mathrm{gm}}(k) + k^2 b_{\nabla^2 \delta} P_{\rm lin}(k) \bigg]}{\mathcal{F} \bigg[ P^{\rm HF}_{\mathrm{mm}}(k) \bigg]} \\
{\rm C} : \frac{\xi_{\mathrm{gm}}}{\xi_{\mathrm{mm}}}&= \frac{\mathcal{F}\bigg[b_1 P^{\rm HF}_{\mathrm{mm}}(k) + P^{\rm 1-Loop}_{\mathrm{gm}}(k) + k^2 b_{\nabla^2 \delta} P^{\rm HF}_{\mathrm{mm}}(k)\bigg]}{\mathcal{F} \bigg[ P^{\rm HF}_{\mathrm{mm}}(k) \bigg]},
\end{aligned}\right.
\end{equation}
where, $\mathcal{F}$ denotes the Fourier transform and $P^{\rm 1-Loop}_{\mathrm{gm}}(k) $ is the effective sum of all the terms dependent on $b_2, b_{\mathrm{s}}$ and $b_{\rm 3nl}$ in Eq.~\ref{eq:P_tm}. An analogous form of this expansion can be derived for $P_{ \mathrm{gg}}(k)$. The term $P^{\rm 1-Loop}_{\mathrm{mm}}(k) $ is the 1-Loop PT estimate of the matter-matter correlation function.  Model A is the linear bias model and the numerator in Model B is similar to the model considered by previous analyses using the EFT description of clustering \citep{Goldberger_06,Senatore_15, Baumann_12,Ivanov_eft:2019,DAmico_eft:2019,perko2016biased,Chudaykin_2019}. In this study, we also analyze Model C, which differs from Model B in the use of the full nonlinear matter power spectrum using {\it halofit} (as opposed to 1-loop PT in Model B) in the numerator. This model is motivated by completely re-summing the matter-matter auto-correlation term to all orders as it uses  the fully non-linear fits to simulations such as {\it halofit}  \citep{Takahashi:2012em}: $P^{\rm NL}_{\mathrm{mm}} = P^{\rm HF}_{\mathrm{mm}}$. We make similar a choice for  $P^{\rm grad}_{\mathrm{mm}}(k)$ \citep{Baldauf15}. The bias term, $b_{\nabla^2 \delta}$ is the sum of both the higher-derivative bias term ($b^{\rm hd}_{\nabla^2 \delta}$)  and the sound speed term ($c^2_{\rm s}$) for $P_{ \mathrm{gm}}(k)$. The sound speed term is zero in Model C as the fully non-linear matter power spectra include any correction from the UV divergent integrals. Hence in Model C, $b_{\nabla^2 \delta} = b^{\rm hd}_{\nabla^2 \delta}$.  Unlike Model C, in Model B the sound speed term is not zero, so there we denote $b_{\nabla^2 \delta} = b^{\rm hd}_{\nabla^2 \delta} + c^2_{\rm s}$. 

The choice of different power spectra for the three models are given in Table~\ref{Tab:models}.

\begin{table}[ht!]
\begin{ruledtabular}
\begin{tabular}{l c c c}
\textrm{Models} & 
\begin{tabular}{@{}c@{}}$P_{\rm mm}$ \end{tabular} &
\begin{tabular}{@{}c@{}}$P^{\rm grad}_{\rm mm}$ \end{tabular} &
\begin{tabular}{@{}c@{}} Remarks \end{tabular}  \\
\hline \hline
Model A  &  $P^{\rm HF}_{\rm mm}$   & 0 & Linear bias model  \\
Model B &  $P^{\rm 1-loop}_{\rm mm}$ & $P_{\rm L}$ & 1-Loop EFT model \\
Model C &  $P^{\rm HF}_{\rm mm}$   &  $P^{\rm HF}_{\rm mm}$   & \textit{Fiducial} model  \\
\end{tabular}
\end{ruledtabular}
\caption{\label{Tab:models}
Variations in the choice of power spectra elements in the three models considered here. Based on the analysis of the three models, we will used Model C as our \textit{fiducial} model (see \S\ref{sec:results})
}
\end{table}

\begin{figure*}[ht]
    \centering
    \includegraphics[width=1.0\textwidth]{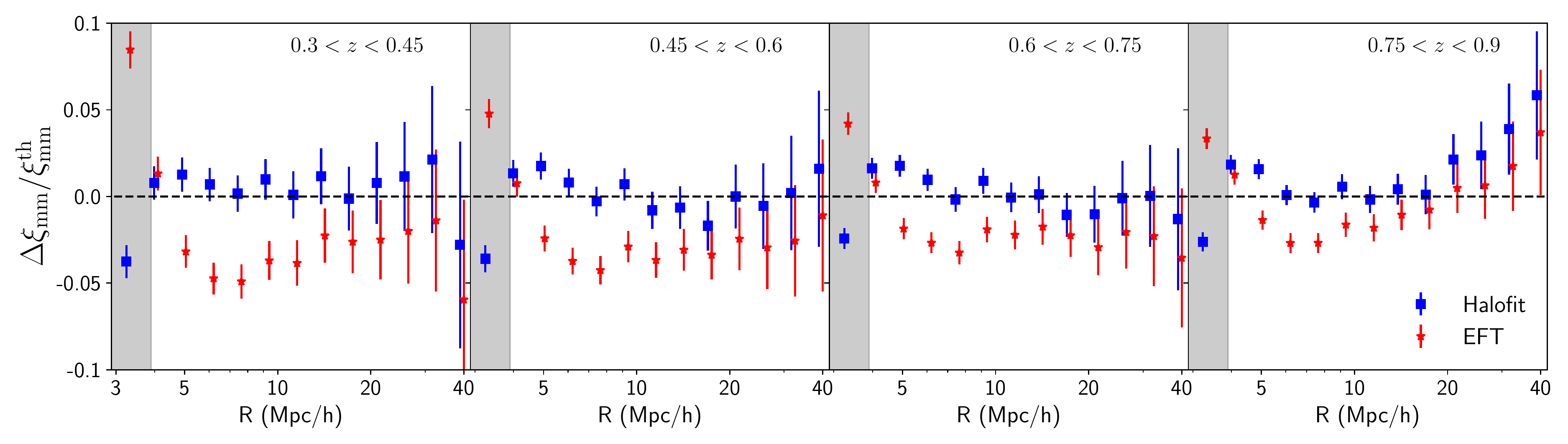}
    \caption{Residuals of the matter-matter correlation function for the four tomographic bins (from left to right) when using {\it halofit} and EFT as the theoretical model. The difference between the model and measurements from the \mice simulations is plotted. {\it Halofit} performs significantly better on small scales. The reduced $\chi^2$ for {\it halofit}  using the data points above 4Mpc/$h$ (outside of the gray shaded regions) are $0.36,0.53,0.49$ and $0.55$ for the four tomographic bins respectively. The red and blue points are staggered for clarity. 
    }
    \label{fig:xi_mm_res}
\end{figure*}

Note that the denominator of Models B and C implicitly assumes that {\it halofit} is a good description of the matter-matter correlation on the scales we are interested in. We check this assumption using the matter density field from the \mice simulations. The residuals of the matter-matter correlation functions for both {\it halofit} and EFT are shown in Fig.~\ref{fig:xi_mm_res}. The EFT theory curve is predicted by fitting the measured $\xi_{\mathrm{mm}}$ on scales larger than 4 Mpc/$h$ with the model: $\xi_{\mathrm{mm}} = \mathcal{F}(P^{\rm 1-Loop}_{\mathrm{mm}}(k) + c^2_{\rm s} k^2 P_{\rm lin}(k))$. We can see that EFT shows deviations at the 5\% level while {\it halofit} is a good description of $\xi_{\mathrm{mm}}$ over all scales and redshifts -- typically within 2\% for the bins with percent level error bars on the measurement.

\subsection{Goodness of fit}
To assess the goodness of fit of the models, we use the reduced $ \chi^2 $. For a good fit to $n_{\mathrm{d}}$ number of data-points, using a model with $n_{\mathrm{v}}$ free parameters, we expect the $\chi^2/{\rm d.o.f}$ to have a mean of $1$ and standard deviation of $\sqrt{2/{\rm d.o.f}}$, where ${\rm d.o.f} = n_{\mathrm{d}} - n_{\mathrm{v}}$ is the total number of degrees of freedom.

\subsection{FAST-PT}
The mode coupling kernels that appear in perturbative terms, such as the higher-order bias contributions in Eq.~\ref{eq:P_tm}, in Fourier space take the form of convolution integrals.
For example in Standard Perturbation Theory, we expand the evolved over-density field of tracers in terms of the linear overdensity, up to third order. This results in terms in the power spectrum that are proportional to $ P_{22}(k) $ (given by the ensemble average $ \langle \delta^{(2)}\delta^{(2)} \rangle $) and $ P_{13}(k) $ (given by $ \langle \delta^{(1)}\delta^{(3)} \rangle $). These kernels can be efficiently evaluated using fast Fourier transform techniques presented in \cite{McEwen2016,Fang2016FAST-PTTheory,Schmittfull_2016}, if one transforms these convolution integrals to the prescribed general form. We use the publicly available Python code \textsc{FAST-PT} as detailed in \citet{McEwen2016} to evaluate all the PT kernels, which is also tested against a C version of the code \textsc{CFASTPT}\footnote{\textsc{FAST-PT} is available at \url{https://github.com/JoeMcEwen/FAST-PT}, and \textsc{CFASTPT} is available at \url{https://github.com/xfangcosmo/cfastpt}}.

\subsection{Covariance Estimation}\label{sec:covaraince}

We estimate a covariance for the data vector by applying the jackknife method \citep{QUENOUILL_56, Tukey_58} to the simulation split into $N_{\rm jk}$ number of patches. We use the k-means clustering algorithm to get the patches, which roughly divides the octant of sky occupied by our galaxy samples into $N_{\rm jk}$ equal-area patches. We use these same patches for covariance calculation in each of our tomographic bins. The accuracy of the estimated covariance increases with increasing $N_{\rm jk}$ and for scales much smaller than the size of an individual patch \citep{Norberg:2008tg,Friedrich:2015nga}. As the total area of the mock catalogs is fixed, changing the number of jackknife patches changes each patch's size.

In order to provide constraints on both non-linear and linear bias parameters, the analysis requires a covariance estimate that correctly captures the auto and cross-correlations between radial bins over both small and large scales to provide constraints on both non-linear and linear bias parameters. We find that we need to limit the analysis to $z > 0.3$ to achieve stable covariance estimates. For this reason, we do not analyze the \mice catalog over the first tomographic bin used in the DES-Y1 analysis ($0.15 < z < 0.3$). 

We estimate the jackknife covariance using $N_{\rm jk} = 300$ patches.  For the lowest redshift bin ($0.3 < z < 0.45$), this results in an individual jackknife patch with a side length of approximately $100$Mpc/$h$. We determine the maximum scale included in our analysis by varying the number of patches and comparing the estimated errors at different scales. 
We find the covariance estimate to be stable below 40 Mpc/$h$ and use this as our maximum scale cut. These tests are detailed in Appendix.~\ref{app:cov}. 

 We explicitly remove the cross-covariance between tomographic bins as there is negligible overlap in the galaxy samples of two different redshift bins, and as length scales of interest are much smaller than the radial extent of the tomographic bins. We correct for biases in the inverse covariance (when calculating the reduced $\chi^2$) due to the finite number of jackknife patches using the procedure described in \citet{Hartlap_2006}.
 
 Note that Fig.~\ref{fig:comp_njk} shows the signal to noise for these 3D statistics for each radial bin for our \textit{fiducial} covariance.

\section{Results}
\label{sec:results}

\subsection{Measurements}\label{sec:measure}
We split the galaxy sample into four tomographic bins, following the DES Year-1 analysis \citet{DESy1}. The redshift ranges for the four bins are: $0.3 < z < 0.45$, $0.45 < z < 0.6$, $0.6 < z < 0.75$ and $0.75 < z < 0.9$ .

\begin{figure*}
    \centering
    \includegraphics[width=1.0\textwidth]{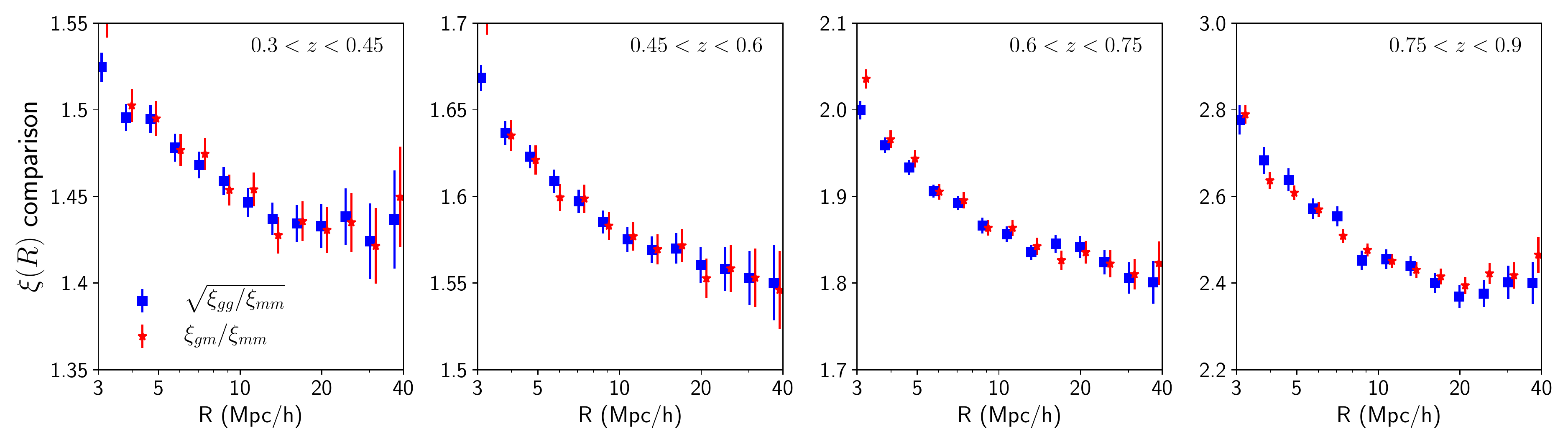}
    \caption{Measurements of ratio of the 3D galaxy-matter correlation functions ($\xi_{\mathrm{gg}}$) and the matter-matter auto correlation ($\xi_{\mathrm{mm}}$) for  the four tomographic bins of the \redmagic galaxy sample in \mice simulations. The errorbars are estimated from jackknife covariances. We fit PT models to the ratios $\xi_{\mathrm{gg}}/\xi_{\mathrm{mm}}$ and $\xi_{\mathrm{gm}}/\xi_{\mathrm{mm}}$ as shown in subsequent figures.}
    \label{fig:xi_gg_gm}
\end{figure*}

The auto and cross-correlations measured with the galaxy and matter catalogs in the \mice simulations are shown in Fig.~\ref{fig:xi_gg_gm}. We use the Landy-Szalay estimator \citep{Landy_Szalay93} to estimate the correlation functions $\xi_{\rm gg}$, $\xi_{\rm gm}$ and $\xi_{\rm mm}$ for all the $N_{\rm jk}$ jackknife patches (see \S\ref{sec:covaraince}). We create a random catalog with 10 times the number of galaxies in each tomographic bin and with number densities corresponding to smoothed galaxy number density. We then use the ratios $\xi_{\rm gg}/\xi_{\rm mm}$ and $\xi_{\rm gm}/\xi_{\rm mm}$ to create our datavector and jackknife covariance. We use the public code Treecorr \cite{Jarvis_2004} to measure the cross correlations. We jointly fit these ratios $\xi_{\mathrm{gg}}/\xi_{\mathrm{mm}}$ and $\xi_{\mathrm{gm}}/\xi_{\mathrm{mm}}$ with PT models mentioned in \S\ref{sec:dv_models}, as described next.

\subsection{Results on fitting the 3D correlation functions}
\label{sec:results3D}
As a first analysis step, we fit the correlation function ratios measured from the simulation with the three models, Model A, B and C (Eq.~\ref{eq:modelABC}) described in \S\ref{sec:dv_models}. Model A only has one free parameter, linear bias $b_1$, while Model B and C in principle have $b_1, b_2, b_{\mathrm{s}}, b_{\rm 3nl}$ and $b_{\nabla^2 \delta}$ as free parameters. Here $b_{\nabla^2 \delta}$ is the higher-derivative bias parameter. Among these parameters, by using the equivalence of Lagrangian and Standard Eulerian perturbation theory (see \S\ref{sec:regPT}), we can write  $b_{\mathrm{s}}$ and $b_{\rm 3nl}$ in terms of $b_1$ as their co-evolution value. Therefore, the simplest complete 1Loop model has $b_1$, $b_2$ and $b_{\nabla^2 \delta}$ as free parameters. We fit our measurements while varying the number of free parameters in both Model B and Model C, to find the minimum number of parameters needed to describe the measured correlation function for different scale cuts.

\begin{figure}
    \centering
    \includegraphics[width=1.0\linewidth]{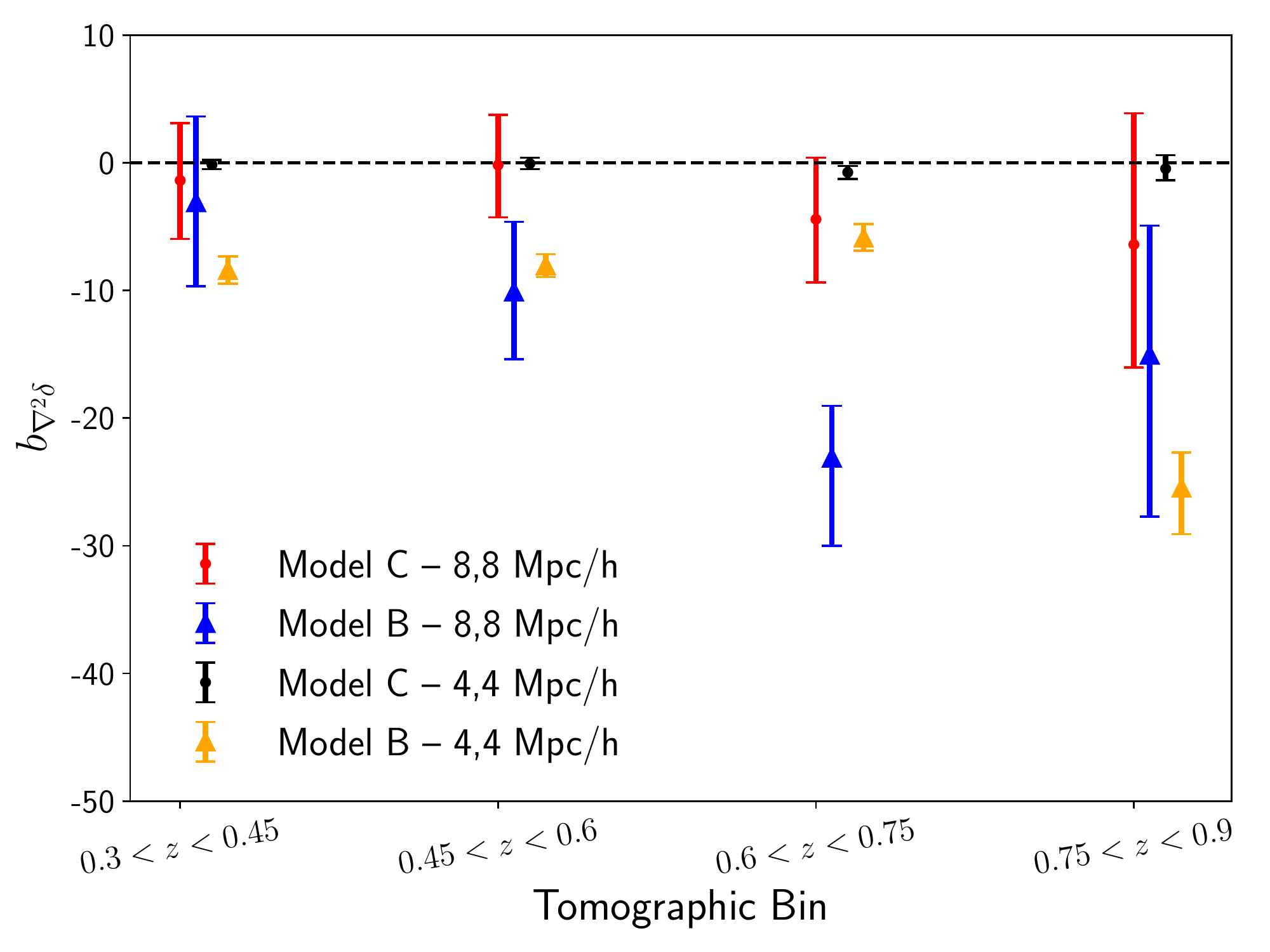}
    \caption{The effective field theory parameter ($b_{\nabla^2 \delta}$) estimated from two different models, described in Eq.~\ref{eq:modelABC}, at two different scale cuts and using the \redmagic galaxy sample. For example, the red points are the result of a joint analysis of ratios $\xi_{\mathrm{gg}}/\xi_{\mathrm{mm}}$ and $\xi_{\mathrm{gm}}/\xi_{\mathrm{mm}}$ (see Fig.~\ref{fig:xi_gg_gm}) above  8Mpc/$h$ using Model C with free $b_1$, $b_2$ and $b_{\nabla^2 \delta}$ parameters for each tomographic bin. We see that when the matter-matter correlation function is described by non-linear {\it halofit} (Model C), the marginalized EFT terms are consistent with zero for all redshifts and both scale cuts. }
    \label{fig:bkcomp}
\end{figure}

We analyze the \mice data-vector with two different minimum scale cuts: 8Mpc/$h$ and 4Mpc/$h$. In Fig.~\ref{fig:bkcomp}, we compare the marginalized constraints on $b_{\nabla^2 \delta}$ for Model B and C for each redshift bin. The marginalized constraints on $b_{\nabla^2 \delta}$ are consistent with zero for Model C, for all redshift bins, and both scale cuts. In contrast, Model B shows significant detection of the $b_{\nabla^2 \delta}$ term. It appears that the EFT term mostly captures the departure of the matter correlation function model from the truth. 

\begin{figure}
    \centering
    \includegraphics[width=1.0\linewidth]{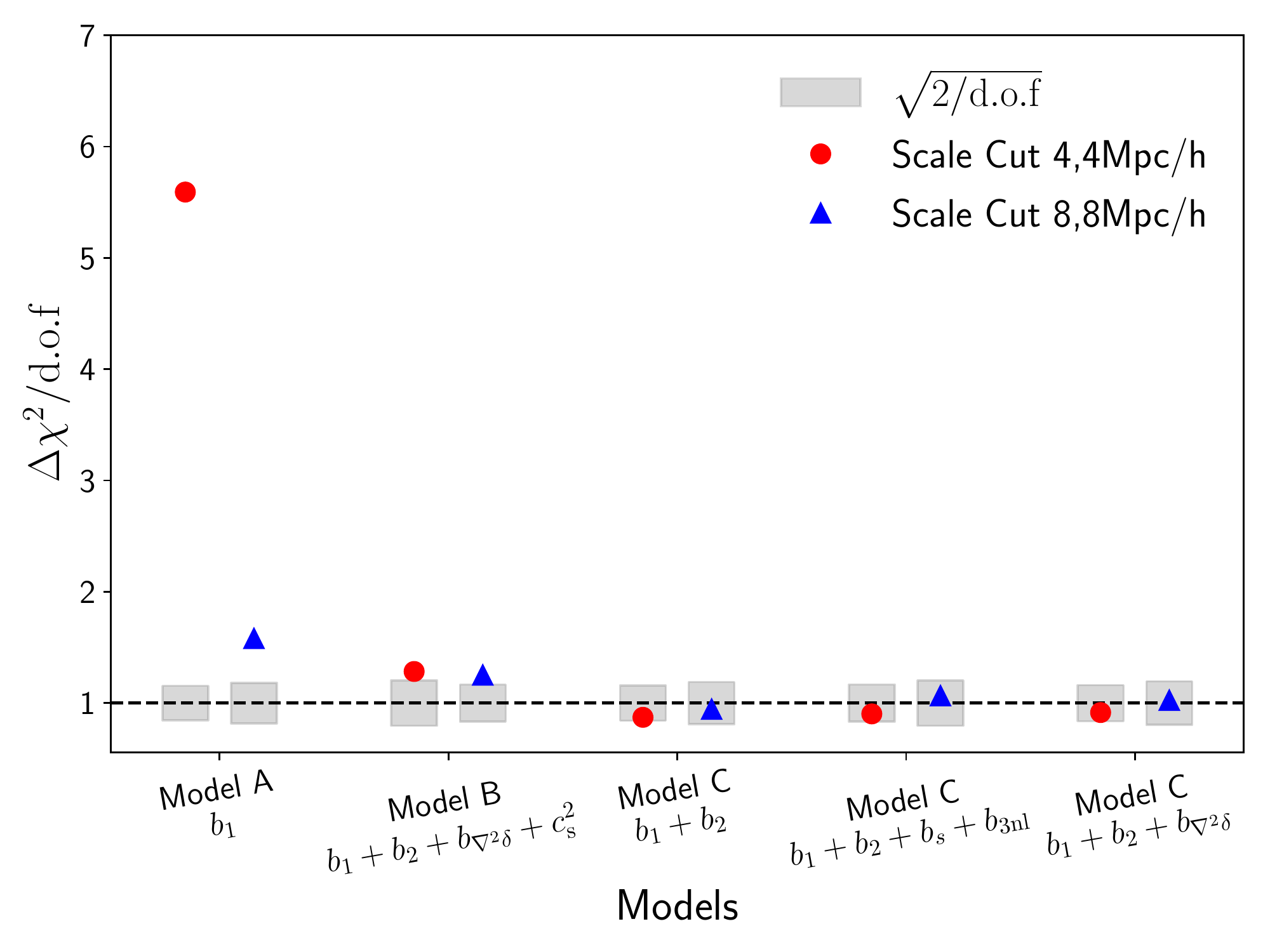}
    \caption{The reduced $\chi^2$ for various choices of free parameters in the models described in Eq.~\ref{eq:modelABC}, when fitting the 3D measurements of the \redmagic galaxy sample at scale cuts of 8Mpc/$h$ and 4Mpc/$h$. The gray band denotes the expected error in the reduced $\chi^2$ for a given number of degrees of freedom. We use  Model C with two free parameters, $b_1$ and $b_2$  as our \textit{fiducial} model (with $b_{\mathrm{s}}$ \& $b_{\rm 3nl}$ fixed to their co-evolution value and $b_{\nabla^2 \delta} = 0$).  }
    \label{fig:redchi2comp}
\end{figure}

\begin{figure*}
    \centering
    \includegraphics[width=1.0\linewidth]{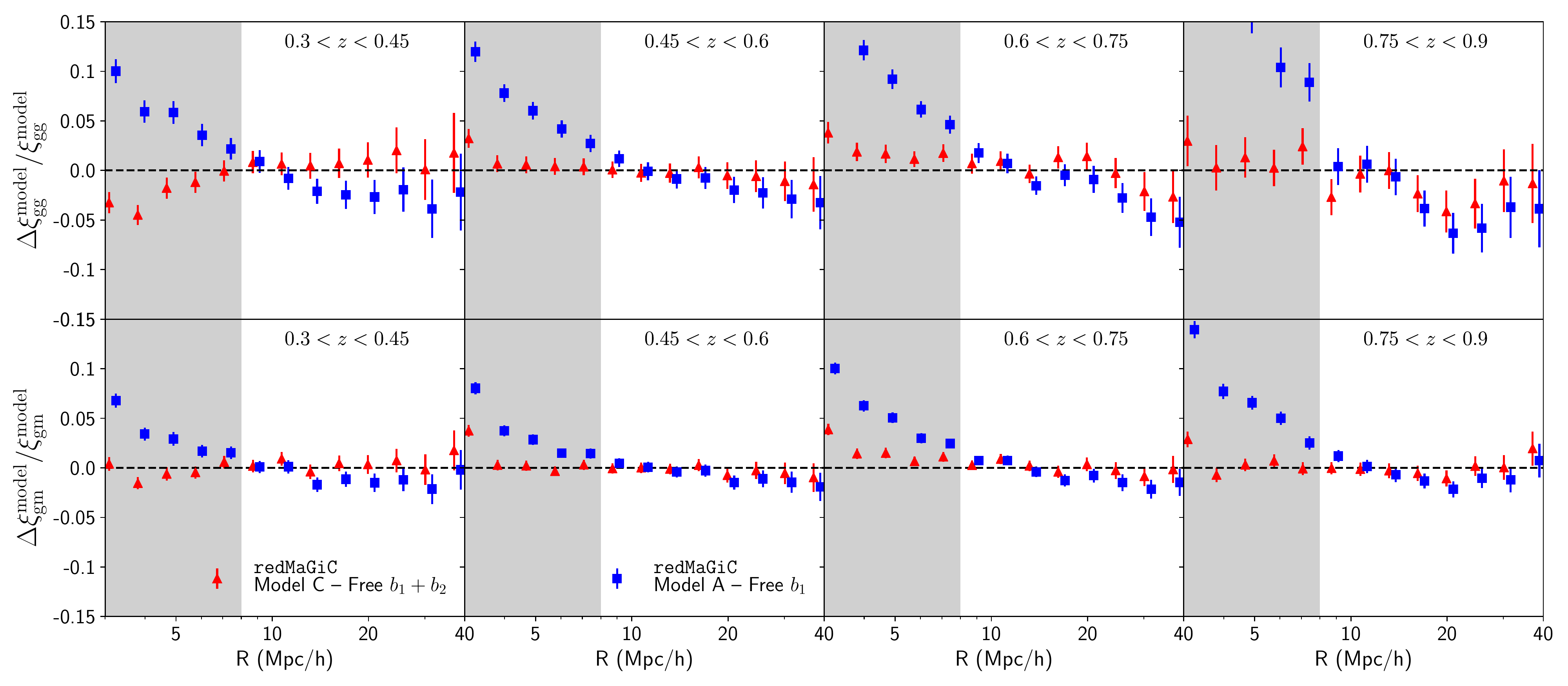}
    \caption{Residuals after doing a joint fit to the measurements of 3D statistics in the \redmagic galaxy sample in four tomographic bins shown in Fig.\ref{fig:xi_gg_gm} with Model A ({\it linear bias model}) and our \textit{fiducial} model, Model C ({\it 1Loop PT} model, with free $b_1$ \& $b_2$ bias parameter for each bin, $b_{\mathrm{s}}$ \& $b_{\rm 3nl}$ fixed to the co-evolution value, $b_{\nabla^2 \delta}=0$) and using {\it halofit} for matter-matter auto-correlation.  Panels in the upper row show the residuals for the galaxy-galaxy correlation function, and panels in the lower row show the residuals for galaxy-matter correlation function. Note that we refer to $\xi^{\rm model}_{\rm gg} = \xi_{\rm gg}/\xi_{\rm mm}$ and $\xi^{\rm model}_{\rm gm} = \xi_{\rm gm}/\xi_{\rm mm}$. Model C is an adequate description of the simulation measurements. We use a scale cut of 8Mpc/$h$ here and only fit the data-points outside the grey region. }
    \label{fig:8mpch}
\end{figure*}

\begin{figure*}
    \centering
\includegraphics[width=1.0\linewidth]{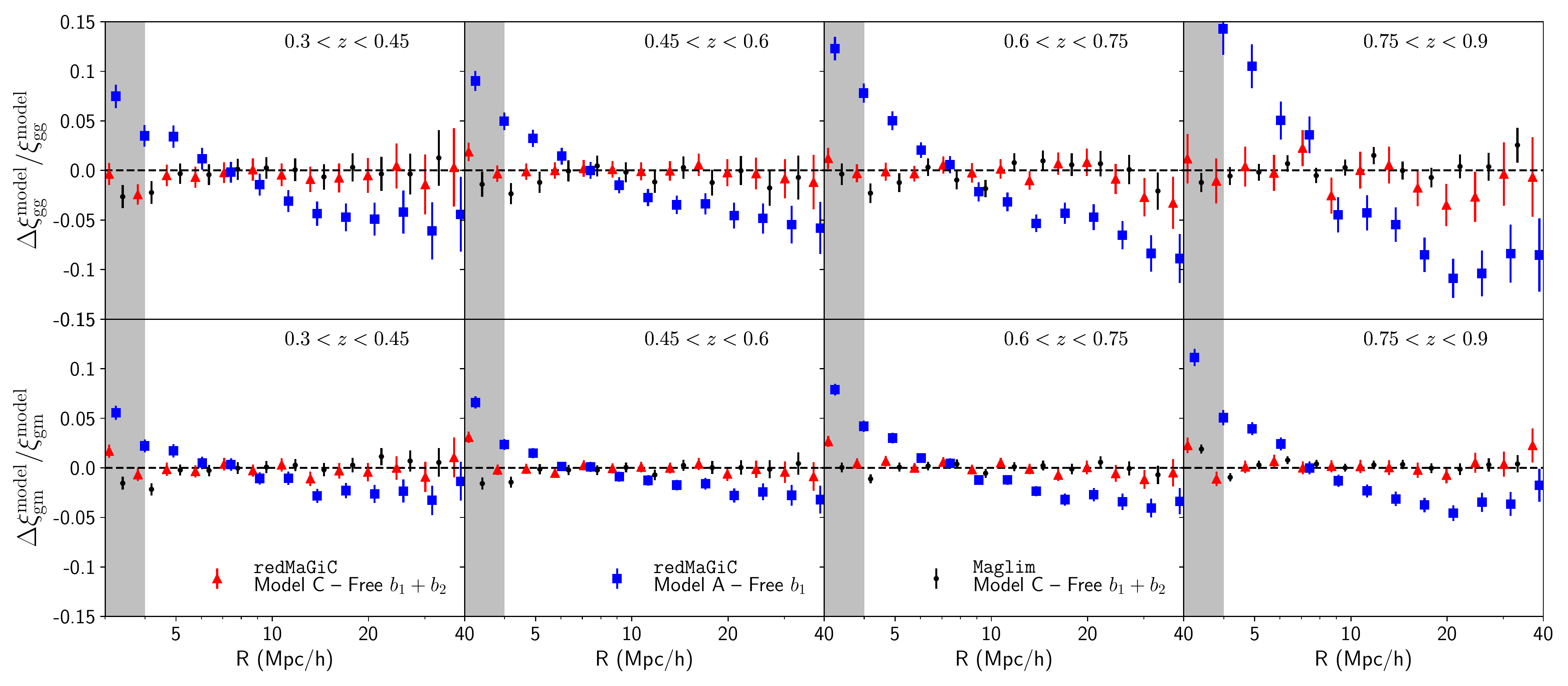}
    \caption{Same as Fig.\ref{fig:8mpch} but analyzed with scale cut of 4Mpc/$h$. Here we also show the residuals for the \maglim galaxy sample. Model C fits the simulation measurements with these smaller scale cuts for both \redmagic and \maglim samples. }
    \label{fig:4mpch}
\end{figure*}

Figure~\ref{fig:redchi2comp} compares the goodness of fit of different models by showing the reduced $\chi^2$ estimated from the best-fit of various model choices (as given in the $x$-axis). We find that using Model C with only $b_1$ and $b_2$ as free parameters gives a reduced $\chi^2$ consistent with 1 for all redshift bins (with $b_{\mathrm{s}}$ \& $b_{\rm 3nl}$ fixed to their co-evolution value and $b_{\nabla^2 \delta}=0$). Hence, we conclude that adding these as free parameters is not needed to model the measurements on the scales considered here. In what follows, we consider this model choice of using 1Loop PT with free $b_1$ and $b_2$ as our \textit{fiducial} model.
We also compare our fits to Model A, with free linear bias parameter $b_1$. The residuals of the observables, i.e., the ratios $\xi_{\mathrm{gg}}/\xi_{\mathrm{mm}}$ and $\xi_{\mathrm{gm}}/\xi_{\mathrm{mm}}$, are shown in Fig.~\ref{fig:8mpch} for a scale cut of 8Mpc/$h$, and in Fig.~\ref{fig:4mpch} for a scale cut of 4Mpc/$h$. Note that  \textit{halofit} describes the matter-matter autocorrelation above scales of 4Mpc/$h$ at about the 2\% level (see Fig.~\ref{fig:xi_mm_res}). In these and following figures, we refer to $\xi^{\rm model}_{\rm gg} = \xi_{\rm gg}/\xi_{\rm mm}$ and $\xi^{\rm model}_{\rm gm} = \xi_{\rm gm}/\xi_{\rm mm}$.
Our \textit{fiducial} model fits the simulations on scales above 4Mpc/$h$ and $z<1$ within 2\%, while the linear bias model performs significantly worse. 

We also show the residuals of our fits to the \maglim sample in Fig.~\ref{fig:4mpch}. We find that similar to the \redmagic sample results, the \textit{fiducial} model describes the measurements within about 2\% above scales of 4Mpc/$h$.

\begin{figure}
    \centering
    \includegraphics[width=1.0\linewidth]{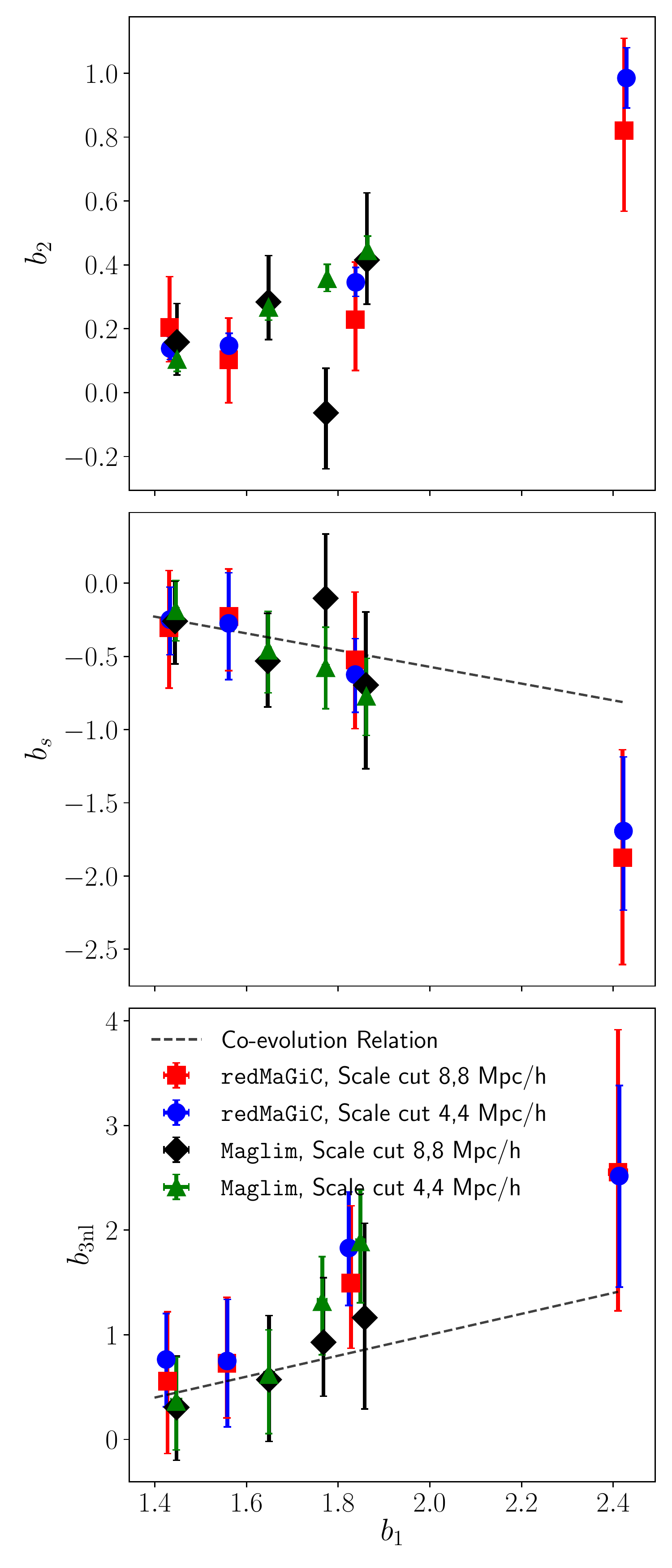}
    \caption{The relation between the best-fit non-linear bias parameters and the linear bias parameter $b_1$ for the four tomographic bin at two different scale cuts. We show the results for both \redmagic and \maglim galaxy samples. The top panel shows second order bias parameter $b_2$ with $b_{\mathrm{s}}$ and $b_{\rm 3nl}$  fixed to their co-evolution Lagrangian values. The middle panel shows  $b_{\mathrm{s}}$  (with $b_{\rm 3nl}$ fixed to the coevolution Lagrangian value). The bottom panel shows $b_{\rm 3nl}$ (with $b_{\mathrm{s}}$ fixed to the coevolution Lagrangian value).  }
    \label{fig:b1bsb3nlcomp}
\end{figure}

\subsection{Relations between bias parameters}

In this section we revisit the approximation that the non-linear bias parameters $b_{\mathrm{s}}$ and $b_{\rm 3nl}$ follow the co-evolution relation. The equivalence of the local Lagrangian and non-local Eulerian description predicts $b_{\mathrm{s}} = -4/7(b_1 - 1)$ and $b_{\rm 3nl} = (b_1 - 1)$ (see \S\ref{sec:regPT}). We test this assumption by freeing up these parameters in addition to $b_1$ and $b_2$ and re-fitting the measurements with these extended models. Figure~\ref{fig:b1bsb3nlcomp} shows the relation between the non-linear bias parameters and $b_1$ at the two scale cuts and for both \redmagic and \maglim galaxy samples. The  points in each panel for each scale cut corresponds to the four tomographic bins. The top panel shows the relation between $b_1$ and $b_2$ (when the parameters $b_{\mathrm{s}}$ and $b_{\rm 3nl}$ are fixed to their co-evolution value), the middle panel shows the relation between $b_1$ and $b_{\mathrm{s}}$ (when $b_{\rm 3nl}$ is fixed to its co-evolution value) and the bottom panel shows the relation between $b_1$ and $b_{\rm 3nl}$ when ($b_{\mathrm{s}}$ is fixed to its co-evolution value). The fits obtained when all the parameters are free have bigger uncertainty but are consistent with the other approaches: the relation between the parameters $b_{\mathrm{s}} - b_1$ and $b_{\rm 3nl} - b_1$ are consistent with the expected co-evolution value. We also note that the recovered relation with $b_1$ is consistent for the two scale cuts, which is a further test that the 1Loop PT is a sufficient and complete model for the scales of interest in this analysis.

It is possible to predict the relation between $b_2$ and $b_1$ for our galaxy samples (the measurements are shown in the top panel of Fig.~\ref{fig:b1bsb3nlcomp}). However, unlike the $b_{\rm s}-b_1$ and $b_{\rm 3nl}-b1$ relation, predicting $b_2-b_1$ relation requires knowledge of the HOD of galaxy samples. Since an accurate HOD of the galaxy sample in data is challenging and not yet available for DES, we have treated $b_2$ as a free parameter. Therefore, only the measurements of the $b_2-b_1$ relation from simulations are shown in Fig.~\ref{fig:b1bsb3nlcomp}.  

\subsection{Inferences for the projected statistics}\label{sec:proj_stats}
\begin{figure}
    \centering
    \includegraphics[width=1.0\linewidth]{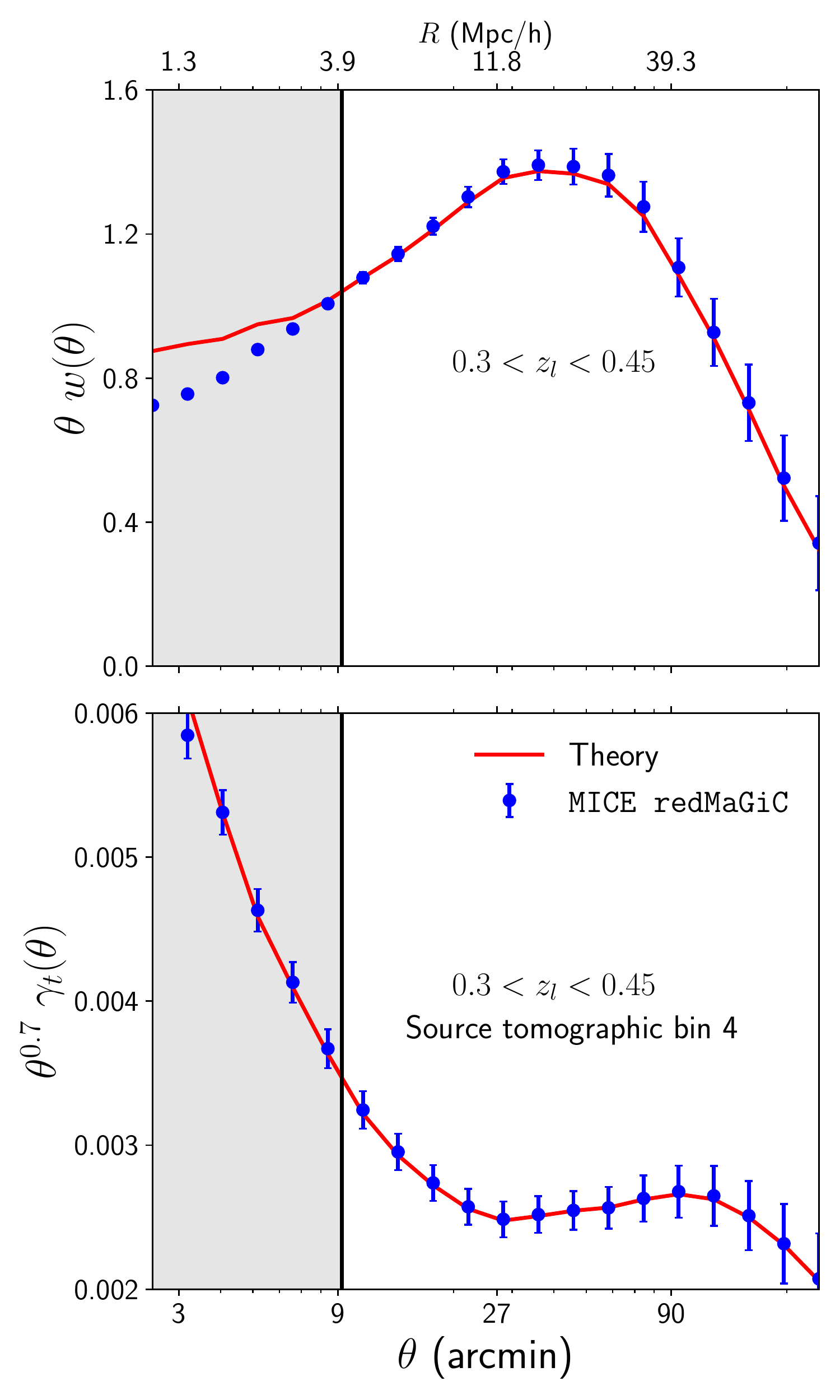}
    \caption{The blue error-bars show the projected statistics $w(\theta)$ and $\gamma_t({\theta})$ transformed from 3D measurements of \mice \redmagic sample (using Eq.~\ref{eq:wgg_t_simp} and Eq.~\ref{eq:gt_b}). The red theory curve is also estimated similarly from the best-fit to the 3D statistics on scales above 4Mpc/$h$ (see Fig.~\ref{fig:4mpch}). These transformations use the $n(z)$ corresponding to our first lens tomographic bin ($0.3 < z_l < 0.45$) (of the \redmagic sample) and fourth source redshift bin (see \cite{DES:Bias}) in the \mice simulation. The shaded region shows the scale cut of 4Mpc/$h$. The errorbars are estimated using the Gaussian halo model covariance for the best-fit bias values. The theory curve for $\gamma_t$ includes the contribution from point-mass (See Eq.~\ref{eq:gt_b}).}
    \label{fig:3d_to_2d}
\end{figure}

As described in \S\ref{sec:3d_to_2d}, we can convert our measurements and fits for the 3D correlation functions to the projected statistics typically used by the imaging surveys.  We show such a conversion in Fig.~\ref{fig:3d_to_2d} for galaxy number densities in \mice simulations corresponding to the \redmagic galaxies satisfying $0.3 < z_{\rm l} < 0.45$ and fourth source tomographic bin as used in the DES Y1 analysis. Note that Fig.~\ref{fig:3d_to_2d} does not show direct measurements of $w(\theta)$ and $\gamma_t$, but a transformation of the measured and best-fit datavector to angular statistics. Since our analysis is based on the ratios $\xi_{\rm gg}/\xi_{\rm mm}$ and $\xi_{\rm gm}/\xi_{\rm mm}$, we first convert our measured datavector and best-fit theory curves to $\xi_{\rm gg}$ and $\xi_{\rm gm}$ and then apply Eq.~\ref{eq:wgg_t_simp} and Eq.~\ref{eq:gt_b} to estimate angular correlation functions. We use {\it halofit} prediction of $\xi_{\rm mm}$, which is a good fit to the  matter-matter autocorrelation for our scales of interest (see Fig.~\ref{fig:xi_mm_res}) to convert the ratios to  $\xi_{\rm gg}$ and $\xi_{\rm gm}$.

The error bars in Fig.~\ref{fig:3d_to_2d} are calculated from Gaussian covariance\footnote{We use the COSMOSIS package \cite{Zuntz_2015} \url{https://bitbucket.org/joezuntz/cosmosis/wiki/Home}} as we do not expect significant non-gaussian contribution to the covariance of the angular statistics (see \citep{Krause:2017}). The covariance is estimated using all the galaxies satisfying the redshift criteria mentioned above in the \mice simulation. Explicitly, we generate this covariance with lens and source galaxies covering 5156.6 square degrees with number densities (per square arc-minutes) of lens galaxies in four tomographic bins corresponding to 0.039, 0.058, 0.045 and 0.028 respectively. The number density and shape noise of source galaxies is assumed to be the same as DES Y3 \citep{Friedrich:inprep}. Due to a similar area and number densities, this covariance is comparable to the expected DES Year-3 covariance \citep{Friedrich:inprep}. Note that the shaded region corresponds to scales below 4Mpc/$h$, which are not used in the 3D fits. The top panel shows the projected galaxy correlation function, $w(\theta)$ and bottom panel shows galaxy-galaxy lensing signal, $\gamma_t(\theta)$.  Note that to estimate $\gamma_t$, we fit for the point-mass term as described in \S\ref{sec:3d_to_2d}. This best-fit value of the point-mass term is obtained by fitting for the coefficient $B$ in Eq.~\ref{eq:gt_b}.

Figure~\ref{fig:3d_to_2d} demonstrates that our model describes the projected angular correlation functions well above scales of 4Mpc/$h$. The error bars in that figure provide a DES Y3 like benchmark for such an agreement. Note that the fractional statistical uncertainties for projected statistics are much larger than their 3D counterparts. Hence the 3D tests presented in \S\ref{sec:results3D} are substantially more stringent than the projected statistics require.

The analysis of measured $w(\theta)$ and $\gamma_t(\theta)$ is detailed in Appendix~\ref{app:2d_stats}.

\subsection{Comparison with other studies in literature}
There have been multiple studies in the literature probing the validity of PT models using simulations \cite{Saito2014a, Angulo_2015, bella2018impact, Werner_2019, alex2020testing}. Most of these studies have focused on Fourier space rather than configuration space. One reason for this choice is that non-linear and linear scales are better separated in Fourier space while in configuration space, even large scales receive a contribution from non-linear Fourier modes. However, many cosmic surveys perform their cosmological parameter analysis in configuration space as it is easier to take into account a non-contiguous mask and depth variations. Hence an understanding of the validity of PT models is required in real space to get unbiased cosmology constraints.

The Fourier space studies conducted by \citet{Saito2014a} and \citet{Angulo_2015} focus only on dark-matter halos and do not aim to reduce the number of free parameters required to explain the auto and cross-correlations between dark matter halos and dark matter particles. \citet{bella2018impact} and \citet{Werner_2019} probe this question on the minimum number of bias parameters but again focus on dark matter halos as the biased tracers. Recently \citet{alex2020testing} have conducted a study similar to ours in Fourier space using three different galaxy samples (mock SDSS and BOSS catalogs) and four halo samples. For a most general case, they find that a four-parameter model (linear, quadratic, cubic non-local bias, and constant shot noise with fixed quadratic tidal bias) can describe correlations between galaxies and matter catalogs, with the inclusion of scale-dependent noise from halo exclusion being particularly beneficial for the combination of auto and cross spectra. They also explore the restriction to a two-parameter model by imposing co-evolution relations, as done in this paper, and find that in general, this reduces the highest Fourier mode for which the model is robust, but it can result in higher constraining power compared to the five parameter model. However, this particular scenario is not general across samples and requires careful validation with simulations, as done here. The main differences in our study are: we work in configuration space with two different galaxy samples that have a higher number density, cover a wider redshift range, and probe smaller host halo masses. Our galaxy samples also have a significantly larger satellite fraction (for example, the first two \redmagic bins have a $\sim 50\%$ satellite fraction) compared to SDSS and BOSS catalogs.

These crucial differences make our study complementary to the above studies. Ours is especially relevant for imaging surveys as it is tailored to DES. The consistency of our conclusions with \citet{alex2020testing} suggests that a two-parameter model may have wide applicability, particularly for surveys with different galaxy selections. This would be an extremely useful result and is worth investigating in detail for the next generation of surveys.

\section{Conclusion}
\label{sec:conclusion}

We have presented an analysis of galaxy bias comparing perturbation theory and 3D correlation functions measured from  N-body simulation-based mock catalogs. We used an effective PT model to analyze the galaxy-galaxy and galaxy-matter correlations jointly. 

Our \textit{fiducial} model successfully describes the measurements from simulations above a scale of 4 Mpc/$h$, which is significantly lower than the scale cut used in the DES Year 1 analysis (where a linear bias model was used). In addition to the linear bias parameter $b_1$, we include four bias parameters in our model: $b_2, b_{\rm s}, b_{\rm 3nl}$ and $b_{\nabla^2 \delta}$. We find that treating only the first and second-order bias parameters $b_1$ and $b_2$ as free parameters is sufficient to describe the correlation functions over the scales of interest. We find that the constraints on the higher-derivative bias parameter $b_{\nabla^2 \delta}$ are consistent with zero in Model C, and we thus fix it to zero in our \textit{fiducial} model. We demonstrate that fixing the parameters $b_{\rm s}$ and $b_{\rm 3nl}$ to their co-evolution value maintains the accuracy of our model. The agreement of our model with measurements from simulations is typically at the 2 percent level over scales of interest. This is within the statistical uncertainty of our simulation measurements and below the requirements of the DES Year 3 analysis. 

We show the relationship between the non-linear and linear bias parameters at different redshifts and scale cuts. We find that the relationship between $b_{\mathrm{s}}-b_1$ and $b_{\rm 3nl}-b_1$ is consistent with the expectations from the co-evolution relationship. Moreover, we find the relationship between $b_2 - b_1$ is consistent at different scale cuts, which is a useful validation of our model. 

We have validated our model with two lens galaxy samples having different and broad host halo mass distribution -- the \redmagic and \maglim samples -- that could be used in DES Y3 cosmological analyses, which combine the projected galaxy clustering signal, $w(\theta)$ and the galaxy-galaxy lensing signal, $\gamma_{\rm t}$. Note that these projected statistics have significantly higher (fractional) cosmic variance than their 3D counterparts $\xi_{\rm gg}$ and $\xi_{\rm gm}$, due to the smaller number of independent modes. Furthermore, the statistical uncertainty of $\gamma_{\rm t}$ includes weak lensing shape noise, which is not included in the error budget of its 3D counterpart ($\xi_{\rm gm}$). Hence, we analyze 3D correlation functions as the measurements from simulations are more precise and provide a percent-level test of our model. 

The scales of interest (above 4 Mpc/$h$) are well above the 1-halo regime, where differences in HOD implementations are greatest. So we expect that our conclusions about bias modeling with PT will have broad validity for the lensing and galaxy clustering analysis from imaging surveys. Nevertheless, at the percent level of accuracy, tests with a variety of schemes for assigning galaxies will be valuable. Moreover, pushing the analysis to higher redshift, or a completely different galaxy selection requires additional testing. We leave these studies for future work. 

\section*{Acknowledgments}

We thank Ravi Sheth for valuable discussions regarding non-linear bias models and the formalism of the paper.

SP and BJ are supported in part by the US Department of Energy Grant No. DE-SC0007901. EK is supported by the US Department of Energy Grant No. DE-SC0020247.

Funding for the DES Projects has been provided by the U.S. Department of Energy, the U.S. National Science Foundation, the Ministry of Science and Education of Spain, 
the Science and Technology Facilities Council of the United Kingdom, the Higher Education Funding Council for England, the National Center for Supercomputing 
Applications at the University of Illinois at Urbana-Champaign, the Kavli Institute of Cosmological Physics at the University of Chicago, 
the Center for Cosmology and Astro-Particle Physics at the Ohio State University,
the Mitchell Institute for Fundamental Physics and Astronomy at Texas A\&M University, Financiadora de Estudos e Projetos, 
Funda{\c c}{\~a}o Carlos Chagas Filho de Amparo {\`a} Pesquisa do Estado do Rio de Janeiro, Conselho Nacional de Desenvolvimento Cient{\'i}fico e Tecnol{\'o}gico and 
the Minist{\'e}rio da Ci{\^e}ncia, Tecnologia e Inova{\c c}{\~a}o, the Deutsche Forschungsgemeinschaft and the Collaborating Institutions in the Dark Energy Survey. 

The Collaborating Institutions are Argonne National Laboratory, the University of California at Santa Cruz, the University of Cambridge, Centro de Investigaciones Energ{\'e}ticas, 
Medioambientales y Tecnol{\'o}gicas-Madrid, the University of Chicago, University College London, the DES-Brazil Consortium, the University of Edinburgh, 
the Eidgen{\"o}ssische Technische Hochschule (ETH) Z{\"u}rich, 
Fermi National Accelerator Laboratory, the University of Illinois at Urbana-Champaign, the Institut de Ci{\`e}ncies de l'Espai (IEEC/CSIC), 
the Institut de F{\'i}sica d'Altes Energies, Lawrence Berkeley National Laboratory, the Ludwig-Maximilians Universit{\"a}t M{\"u}nchen and the associated Excellence Cluster Universe, 
the University of Michigan, the National Optical Astronomy Observatory, the University of Nottingham, The Ohio State University, the University of Pennsylvania, the University of Portsmouth, 
SLAC National Accelerator Laboratory, Stanford University, the University of Sussex, Texas A\&M University, and the OzDES Membership Consortium.

Based in part on observations at Cerro Tololo Inter-American Observatory, National Optical Astronomy Observatory, which is operated by the Association of 
Universities for Research in Astronomy (AURA) under a cooperative agreement with the National Science Foundation.

The DES data management system is supported by the National Science Foundation under Grant Numbers AST-1138766 and AST-1536171.
The DES participants from Spanish institutions are partially supported by MINECO under grants AYA2015-71825, ESP2015-66861, FPA2015-68048, SEV-2016-0588, SEV-2016-0597, and MDM-2015-0509, 
some of which include ERDF funds from the European Union. IFAE is partially funded by the CERCA program of the Generalitat de Catalunya.
Research leading to these results has received funding from the European Research
Council under the European Union's Seventh Framework Program (FP7/2007-2013) including ERC grant agreements 240672, 291329, and 306478.
We  acknowledge support from the Brazilian Instituto Nacional de Ci\^encia
e Tecnologia (INCT) e-Universe (CNPq grant 465376/2014-2).

This manuscript has been authored by Fermi Research Alliance, LLC under Contract No. DE-AC02-07CH11359 with the U.S. Department of Energy, Office of Science, Office of High Energy Physics. The United States Government retains and the publisher, by accepting the article for publication, acknowledges that the United States Government retains a non-exclusive, paid-up, irrevocable, world-wide license to publish or reproduce the published form of this manuscript, or allow others to do so, for United States Government purposes.

\clearpage 

\appendix

\section{Covariance of the data-vectors}\label{app:cov}
\begin{figure}
    \centering
    \includegraphics[width=1.0\linewidth]{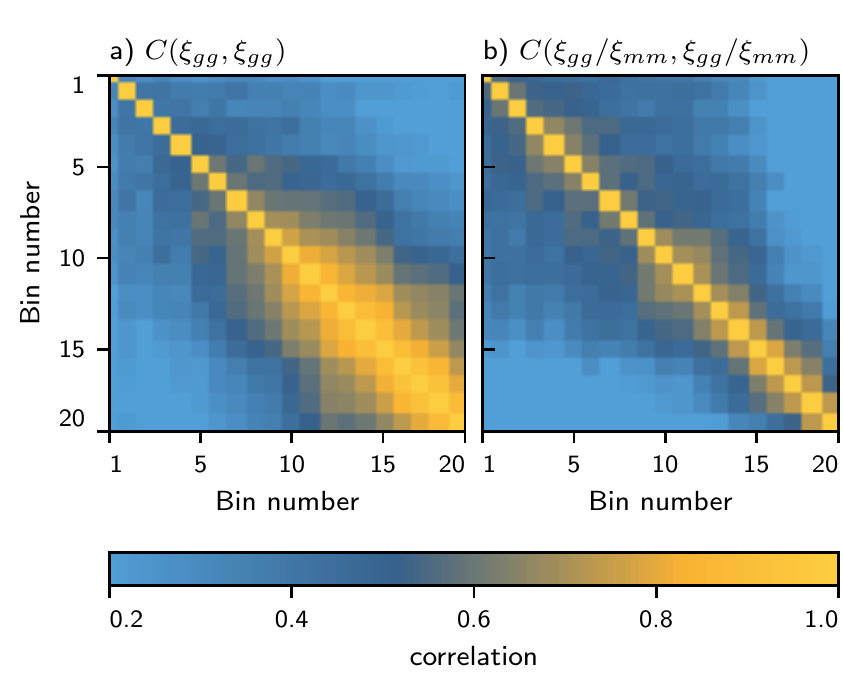}
    \caption{The correlation matrix for the two-point galaxy correlation function $\xi_{\mathrm{gg}}$ and the ratio $\xi_{\mathrm{gg}}/\xi_{\mathrm{mm}}$ for the second tomographic bin. Both correlation matrices are estimated using 300 jackknife patches. We see that the covariance is more diagonal for the ratio. }
    \label{fig:corr_bin3}
\end{figure}

The measurements of the correlation functions  $\xi_{\mathrm{gg}}$ and $\xi_{\mathrm{gm}}$ are highly correlated in the configuration space due to the mixing of modes. However, since the correlation function $\xi_{\mathrm{mm}}$ is also impacted by similar mode-mixing, analyzing the ratio of the correlation functions $\xi_{\mathrm{gg}}/\xi_{\mathrm{mm}}$ and $\xi_{\mathrm{gm}}/\xi_{\mathrm{mm}}$ makes the covariance more diagonal. In the Fig.~\ref{fig:corr_bin3} we compare the correlation matrix for $\xi_{\mathrm{gg}}$ and $\xi_{\mathrm{gg}}/\xi_{\mathrm{mm}}$ for the third tomographic bin for 20 radial bins ranging from 0.8-50 Mpc/$h$. We clearly see that analyzing the ratio gives us much better behaved correlation matrix. 

\begin{figure}
    \centering
    \includegraphics[width=1.0\linewidth]{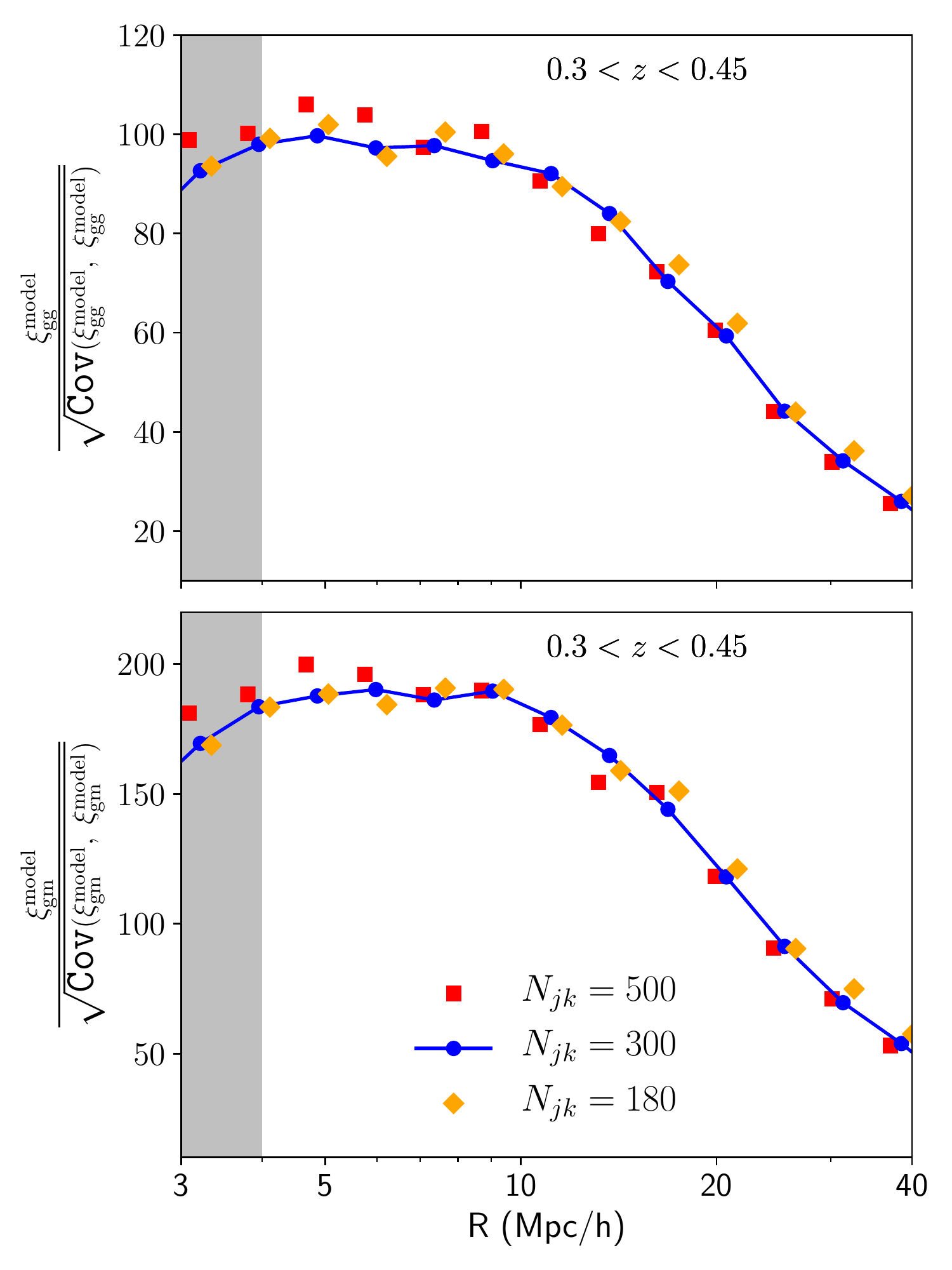}
    \caption{The comparison of errorbars (and signal to noise) estimated using the jackknife procedure, for a different number of patches. We show the comparison for the smallest tomographic bin used in our analysis since that is most susceptible to the sizes of the jackknife patches. Also, since the covariance matrix of the correlation function ratios has small cross-bin covariance (see Fig.~\ref{fig:corr_bin3}), we only compare the diagonal value. The blue points (and solid) curve corresponds to our \textit{fiducial} choice of 300 as the number of jackknife patches used for covariance estimation. }
    \label{fig:comp_njk}
\end{figure}

We generate the \textit{fiducial} jackknife covariance from 300 patches distributed over the simulation footprint. As the total area populated by both our galaxy sample is equal to one octant of the sky, changing the number of jackknife patches, changes the size of each patch.  In Fig.\ref{fig:comp_njk}, we compare the signal to noise estimate when using a different number of patches. We see that the diagonal elements of the covariance are robust to changes in the number of patches. We have also compared the changes in best-fit curves when using the covariance matrix generated using a different number of patches. We get consistent reduced $\chi^2$ and best-fit curves for $z > 0.3$. However, we find that we can not get a robust covariance for the tomographic bin corresponding to $z < 0.3$ without sacrificing large scale information (which is required to constrain the linear bias parameter). For this reason, we only analyze the tomographic bins satisfying $z > 0.3$ and find that with 300 patches, we can get a robust estimate of jackknife covariance.

\section{Results with fitting $\xi_{\mathrm{gg}}$ and $\xi_{\mathrm{gm}}$ directly }\label{app:fit_corr}

\begin{figure}
    \centering
    \includegraphics[width=1.0\linewidth]{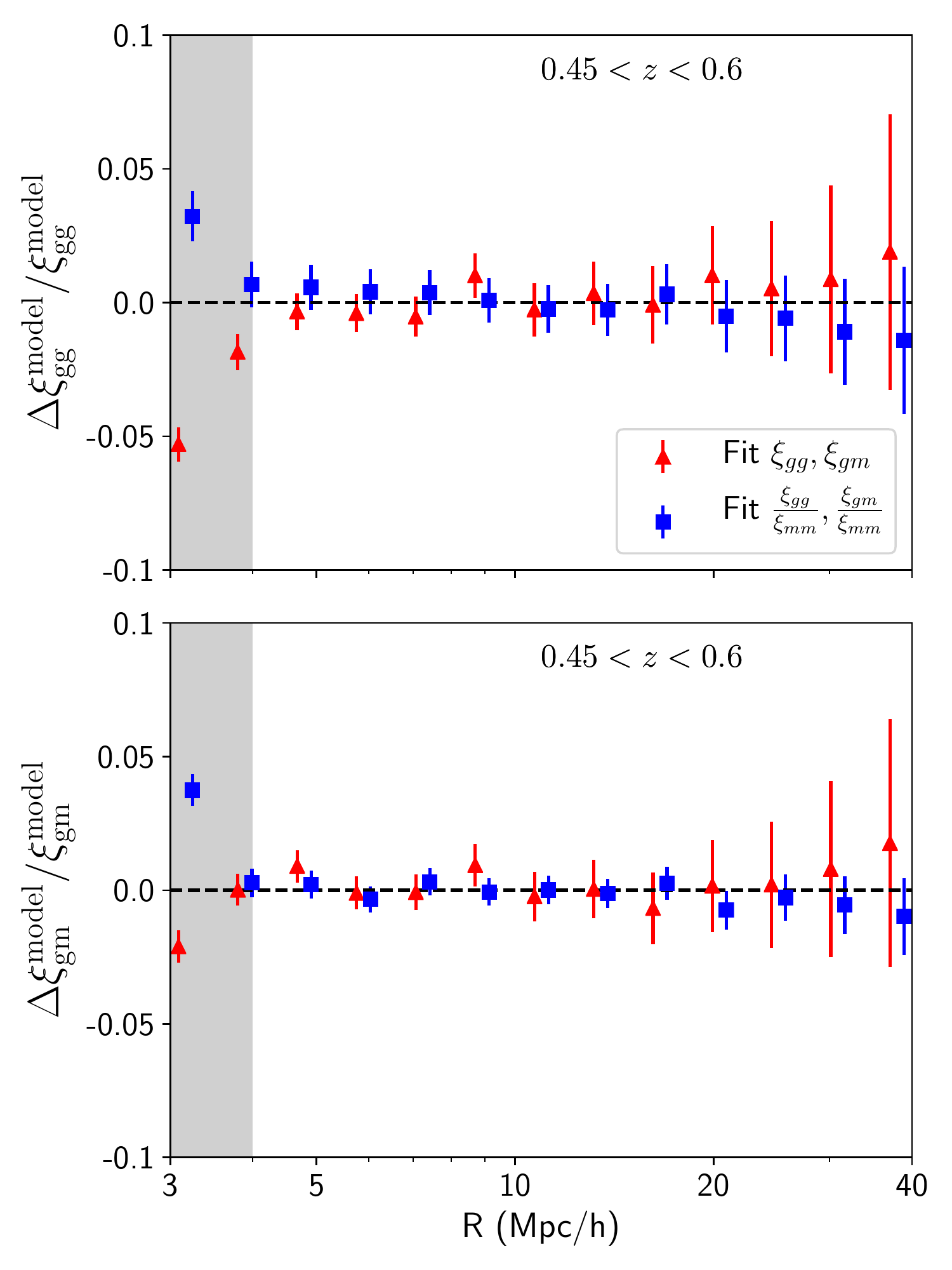}
    \caption{Comparing the residuals when fitting the measured correlation functions $\xi_{\mathrm{gg}},\xi_{\mathrm{gm}}$ directly and when fitting the ratio $\xi_{\mathrm{gg}}/\xi_{\mathrm{mm}},\xi_{\mathrm{gm}}/\xi_{\mathrm{mm}}$ for the second tomographic bin. We use our \textit{fiducial} model as our theory model in both cases. We find the fits are consistent. }
    \label{fig:gg_mm_comp_bin3}
\end{figure}

As mentioned in the main text, we consider the ratios, $\xi_{\mathrm{gg}}/\xi_{\mathrm{mm}}$ and $\xi_{\mathrm{gm}}/\xi_{\mathrm{mm}}$, as our data-vector. This ratio is more sensitive to the galaxy-matter connection than the correlation functions  $\xi_{\mathrm{gg}}$ and $\xi_{\mathrm{gm}}$ themselves. However, when we try to fit directly the correlation functions, $\xi_{\mathrm{gg}}$ and $\xi_{\mathrm{gm}}$, our conclusions do not change. The residuals of the $\xi_{\mathrm{gg}}$ and $\xi_{\mathrm{gm}}$ using our \textit{fiducial} model are shown in Fig.~\ref{fig:gg_mm_comp_bin3} for the third tomographic bin. We compare the residuals obtained when directly fitting the correlation functions $\xi_{\mathrm{gg}}$, $\xi_{\mathrm{gm}}$  with the results shown in the main text obtained when fitting the ratios of the correlation functions, $\xi_{\mathrm{gg}}/\xi_{\mathrm{mm}},\xi_{\mathrm{gm}}/\xi_{\mathrm{mm}}$. We find that our residuals are consistent with zero above the scales of 4Mpc/$h$ for both data-vectors. 

\section{Analyzing the 2D correlation function at fixed cosmology}\label{app:2d_stats}
As described in the section \S\ref{sec:3d_to_2d} and Fig.~\ref{fig:3d_to_2d}, we convert the 3D statistics to the projected statistics. However, we can also fit our perturbation theory models directly to the measured projected statistics. Therefore, in this appendix, we fit our \textit{fiducial} model to the projected statistics $w(\theta)$ and $\gamma_t$ in the four lens and source tomographic bins. We refer the readers to \citet{DES:Bias} for the details about the estimation of the projected statistics and the tomographic redshift distribution of our bins.   

\begin{figure*}
    \centering
    \includegraphics[width=1.0\linewidth]{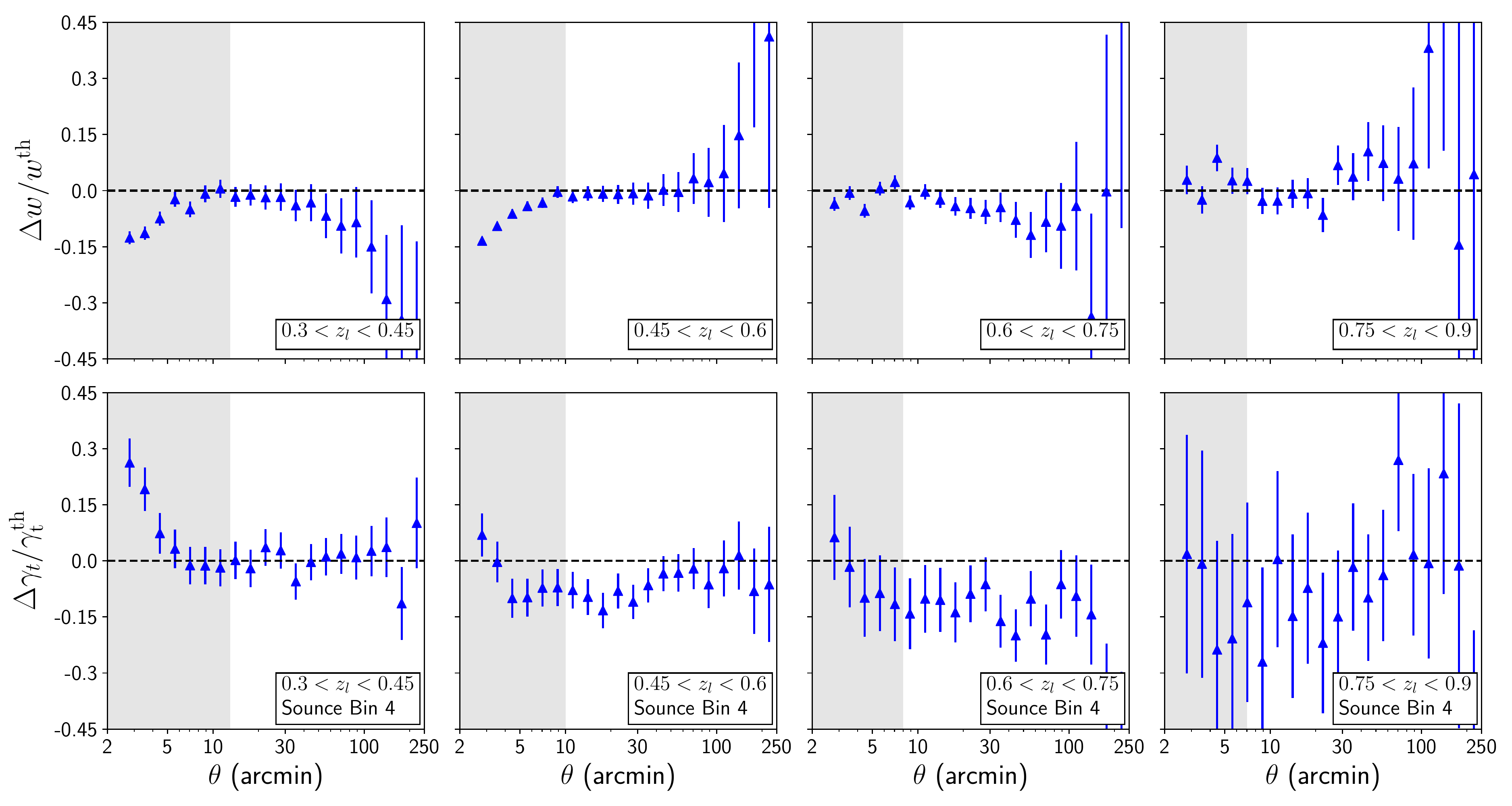}
    \caption{Residual from joint fits to the measurements of the 2D statistics,  in four tomographic lens and source bins (see \cite{DES:Bias} for source $n(z)$), using our \textit{fiducial} model. The top row show the residuals for  $w(\theta)$, and the bottom row for $\gamma_t$, with the source redshift distribution taken as the fourth bin in the DES Y1 analysis. We use a scale cut of 4Mpc/$h$ here and only fit the data-points outside the grey region. The reduced $\chi^2$ including all the datapoints (total degrees of freedom=342) above the scale cut is 0.88. }
    \label{fig:wtgt_4mpch}
\end{figure*}

\begin{figure}
    \centering
    \includegraphics[width=1.0\linewidth]{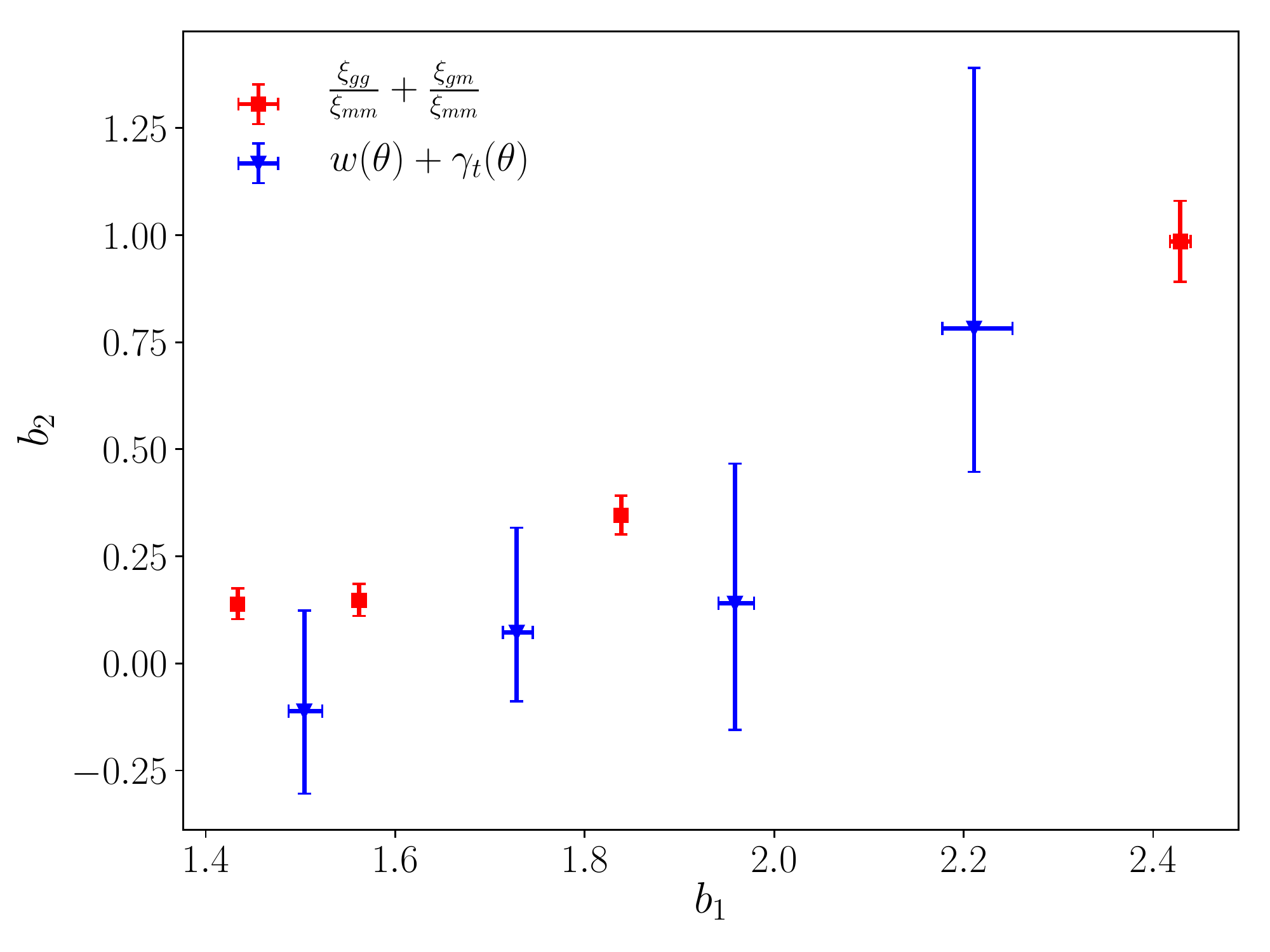}
    \caption{The relation between the marginalized non-linear and linear bias parameters for four tomographic bins estimated from fitting $w(\theta)$ and $\gamma_t$. We also compare these values to the ones estimated from the 3D correlation functions and find consistent $b_2 - b_1$ relation.  }
    \label{fig:b2b1_wtgt}
\end{figure}

The residuals of this model are shown in Fig.~\ref{fig:wtgt_4mpch} when using scales above 4Mpc/$h$. For the observable $\gamma_t$, we show the results for only the fourth source bin and all four lens tomographic bin (since this has the highest signal to noise). The fit has a reduced $\chi^2$ of 0.88. There are some points in the residuals that are inconsistent with zero; however, as there is a significant correlation between different radial bins, they do not impact the $\chi^2$ of the fit. The measured relation between $b_2$ and $b_1$ from this model is shown in Fig.~\ref{fig:b2b1_wtgt}. We also compare this relationship with the one inferred from the 3D measurements and find them consistent.

Hence, when fitting the measured projected correlation functions directly, we also get a reduced $\chi^2$ consistent with one. These results motivate us to model the correlations on the scales down to 4 Mpc/$h$ in the DES Y3 cosmological analysis. To determine the scale cuts for DES analysis with non-linear bias model, we will study the cosmological parameter biases in a future study with the range of scale cut choices motivated by this study.

\bibliography{ref}

\end{document}